\documentclass[journal, onecolumn, 12pt, draftcls]{IEEEtran}

\usepackage{cite}
\usepackage{enumerate}
\usepackage{graphicx}
\usepackage[cmex10]{amsmath} 
\usepackage{amssymb}
\usepackage{mathtools}
\usepackage{xspace}
\usepackage{subfigure}
\usepackage{siunitx}

\usepackage{tikz}
\usepackage{pgfplots}
\usetikzlibrary{arrows,shapes,backgrounds}
\pgfplotsset{compat=newest}                         
\pgfplotsset{plot coordinates/math parser=false}
\newlength\figureheight
\newlength\figurewidth

\DeclareMathOperator{\erfc}{erfc}
\DeclareMathOperator{\err}{err}

\newtheorem{theorem}{Theorem}[section]
\newtheorem{lemma}[theorem]{Lemma}
\newtheorem{definition}[theorem]{Definition}
\newtheorem{corollary}[theorem]{Corollary}
\newtheorem{proposition}[theorem]{Proposition}

\newcommand{\Z}{\mathbb{Z}}
\newcommand{\R}{\mathbb{R}}
\newcommand{\SNR}{\Gamma}
\newcommand{\snr}{\gamma}
\newcommand{\AVESNR}{\overline{\gamma}}

\newcommand{\IND}{\mathcal{I}}
\newcommand{\IID}{\emph{i.i.d.}}
\newcommand{\Fit}{f_i^{(t)}}
\newcommand{\A}{\mathcal{A}}
\newcommand{\F}{\mathcal{F}}
\newcommand{\G}{\mathcal{G}}   
\newcommand{\B}{\mathcal{B}} 

\newcommand{\LN}{L^{\natural}}
\newcommand{\One}{\mathbf{1}}
\newcommand{\Thv}{\mbox{\boldmath$\phi$}}  

\newcommand{\X}{\mathcal{X}}
\newcommand{\FDef}{$\mathcal{F}_i=\{0, 1\}$}
\newcommand{\x}{\mathbf{x}}
\newcommand{\y}{\mathbf{y}}
\newcommand{\act}{\mathbf{a}}
\newcommand{\bv}{\mathbf{b}}
\newcommand{\gv}{\mathbf{g}}
\newcommand{\fv}{\mathbf{f}}
\newcommand{\Pxx}{P_{\x\x^\prime}^\act}

\newcommand{\Pbbi}{P_{b_ib_i^\prime}^{a_i}}
\newcommand{\Pggi}{P_{g_ig_i^\prime}}
\newcommand{\Pbbv}{P_{\bv\bv'}^{\act}}
\newcommand{\Pggv}{P_{\gv\gv'}}

\newcommand{\ev}{\mathbf{e}}

\newcommand{\Pro}{\mathcal{P}^{\star}\!}   

\newcommand{\E}{\mathbb{E}}

\begin{document}

\title{Structured Optimal Transmission Control in Network-coded Two-way Relay Channels}


\author{
Ni~Ding,~\IEEEmembership{Student Member,~IEEE}, Parastoo~Sadeghi,~\IEEEmembership{Senior Member,~IEEE}, Rodney~A.~Kennedy,~\IEEEmembership{Fellow,~IEEE} \\
\thanks{
*Ni Ding is with the Research School of
Engineering, College of Engineering and Computer Science, the Australian National University (ANU), Canberra, ACT 0200, Australia (email: $\{$ni.ding$\}$@anu.edu.au).

Parastoo~Sadeghi and Rodney~A.~Kennedy are with the Research School of
Engineering, College of Engineering and Computer Science, the Australian National University (ANU), Canberra, ACT 0200, Australia (email: $\{$parastoo.sadeghi, rodney.kennedy$\}$@anu.edu.au).
}
}

\markboth{Structured Optimal Transmission Control in Network-coded Two-way Relay Channels}%
{Ding, Sadeghi and Kennedy}

\maketitle

\begin{abstract}
This paper considers a transmission control problem in network-coded two-way relay channels (NC-TWRC), where the relay buffers random symbol arrivals from two users, and the channels are assumed to be fading. The problem is modeled by a discounted infinite horizon Markov decision process (MDP). The objective is to find a transmission control policy that minimizes the symbol delay, buffer overflow and transmission power consumption and error rate simultaneously and in the long run. By using the concepts of submodularity, multimodularity and $\LN$-convexity, we study the structure of the optimal policy searched by dynamic programming (DP) algorithm. We show that the optimal transmission policy is nondecreasing in queue occupancies or/and channel states under certain conditions such as the chosen values of parameters in the MDP model, channel modeling method, modulation scheme and the preservation of stochastic dominance in the transitions of system states. The results derived in this paper can be used to relieve the high complexity of DP and facilitate real-time control.
\end{abstract}

\begin{IEEEkeywords}
discounted Markov decision process, dynamic programming, $\LN$-convexity, monotonic optimal policy, multimodularity, network coding, submodularity, supermodular game.
\end{IEEEkeywords}


\IEEEpeerreviewmaketitle

\section{Introduction}

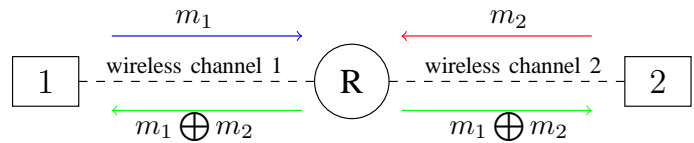
\begin{figure}[h]
	\centering
		\centerline{\scalebox{1.1}{\begin{tikzpicture}

\draw (-1.6,0.3) rectangle (-0.8,-0.3);
\node at (-1.2,0) {$1$};
\draw [dashed] (-0.8,0) -- (2.05,0);

\draw (2.5,0) circle (0.45);
\node at (2.5,0) {R};

\draw (5.8,0.3) rectangle (6.6,-0.3);
\node at (6.2,0) {$2$};
\draw [dashed] (2.95,0) -- (5.8,0);

\draw [ ->,color=blue] (-0.4,0.55) -- (1.9,0.55);
\draw [->,color=green] (3.1,-0.35) -- (5.4,-0.35);
\node at (0.6,0.75) {\small $m_1$};

\draw [<-,color=red] (3.1,0.55) -- (5.4,0.55);
\draw [<-,color=green] (-0.4,-0.35) -- (1.9,-0.35);
\node at (4.4,0.75) {\small $m_2$};

\node at (0.6,-0.55) {\small $m_1\bigoplus{m_2}$};
\node at (4.4,-0.55) {\small $m_1\bigoplus{m_2}$};

\node at (0.6,0.2) {\scriptsize wireless channel 1};
\node at (4.45,0.2) {\scriptsize wireless channel 2};

\end{tikzpicture}}}
	\caption{NC-TWRC \cite{Hausl2006}. Two users exchange information ($m_1$ and $m_2$) via the center node $R$ (stands for relay).}
	\label{fig:3Node}
\end{figure}

Network coding (NC) was proposed in \cite{Ahlswede2000} to maximize the information flow in a wired network. It was introduced in multicast wireless communications to optimize the throughput, and has gained a lot of  interest recently due to the rapid growth in multimedia applications \cite{WuXOR2004}. It was shown in \cite{KattiXOR2006} that the power efficiency in wireless transmission systems could be improved by NC. For example, in a 3-node network system, called the network-coded two-way relay channels (NC-TWRC)\cite{Hausl2006} as shown in Fig.~\ref{fig:3Node}, the messages $m_1$ and $m_2$ are XORed at the relay and broadcast to the end users. This method, compared to the conventional store-and-forward transmission, reduces the total number of transmissions from $4$ to $3$ so that the transmission power is saved by $25\%$. Since then, numerous optimization problems have been studied in NC-TWRC, e.g., training design for optimal channel estimation \cite{Gao2009} and the design principle of modulation and NC scheme\cite{Koike2009}.

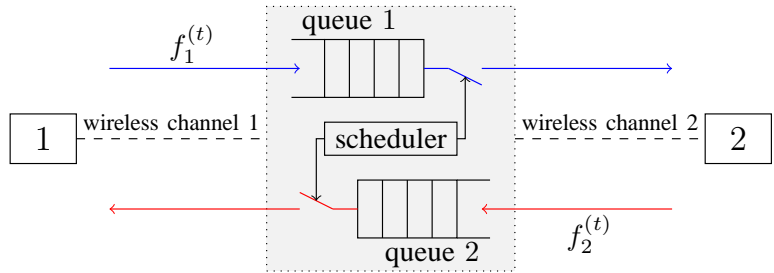
\begin{figure}[h]
	\centering
		\centerline{\scalebox{1.1}{\begin{tikzpicture}

\draw (-1.6,0.3) rectangle (-0.8,-0.3);
\node at (-1.2,0) {$1$};
\draw [dashed] (-0.8,0) -- (1.5,0);

\draw (6.8,0.3) rectangle (7.6,-0.3);
\node at (7.2,0) {$2$};
\draw [dashed] (4.5,0) -- (6.8,0);

\draw [dotted, fill=gray!10] (1.5,1.6) rectangle (4.5,-1.6);
\node at (3,0) {\small scheduler};
\draw (2.2,0.2) rectangle (3.8,-0.2);

\draw (2.2,1.2) rectangle (3.4,0.5);
\draw (2.2,1.2) -- (1.8,1.2);
\draw (2.2,0.5) -- (1.8,0.5);
\draw (2.5,1.2) -- (2.5,0.5);
\draw (2.8,1.2) -- (2.8,0.5);
\draw (3.1,1.2) -- (3.1,0.5);
\draw [ ->,color=blue] (-0.4,0.85) -- (1.9,0.85);
\draw [color=blue] (3.4,0.85) -- (3.7,0.85);
\draw [color=blue] (3.7,0.85) -- (4.1,0.65);
\draw [->,color=blue] (4.1,0.85) -- (6.4,0.85);
\node at (2.5,1.4) {\small queue $1$};
\node at (0.6,1.15) {\small $f_1^{(t)}$};
\draw [->] (3.8,0) -- (3.9,0) -- (3.9,0.75);

\draw (3.8,-1.2) rectangle (2.6,-0.5);
\draw (3.8,-1.2) -- (4.2,-1.2);
\draw (3.8,-0.5) -- (4.2,-0.5);
\draw (3.5,-1.2) -- (3.5,-0.5);
\draw (3.2,-1.2) -- (3.2,-0.5);
\draw (2.9,-1.2) -- (2.9,-0.5);
\draw [<-,color=red] (4.1,-0.85) -- (6.4,-0.85);
\draw [color=red] (2.6,-0.85) -- (2.3,-0.85);
\draw [color=red] (2.3,-0.85) -- (1.9,-0.65);
\draw [<-,color=red] (-0.4,-0.85) -- (1.9,-0.85);
\node at (3.5,-1.4) {\small queue $2$};
\node at (5.4,-1.15) {\small $f_2^{(t)}$};
\draw [->] (2.2,0) -- (2.1,0) -- (2.1,-0.75);

\node at (0.35,0.2) {\scriptsize wireless channel 1};
\node at (5.65,0.2) {\scriptsize wireless channel 2};

\end{tikzpicture}}}
	\caption{NC-TWRC with random symbol arrivals and fading channels \cite{Ding2012}. The stochastic inflows are buffered by two finite length first-in-first-out (FIFO) queues. The outflows are controlled by a scheduler. }
	\label{fig:2U1R_M_W}
\end{figure}

In \cite{KattiImportance2005}, Katti et al.\ pointed out the importance of being opportunistic in practical NC scenarios. It was suggested that the assumptions in the related research work should comply with the practical wireless environments, e.g., decentralized routing and non-stationary traffic rate. This suggestion highlighted a problem in the existing literature: The majority of the studies consider synchronized traffic while ignoring the stochastic nature of the packet arrivals in the network layer. However, introducing the randomness of traffic in Fig.~\ref{fig:3Node} gives rise to a power-delay tradeoff: When there are symbol inflows in the relay but no coding opportunities or XORing pairs (e.g., one symbol arrives from one user, but no symbol arrives from the other), waiting for coding opportunities by holding symbols saves transmission power but increases symbol delay. This dilemma was studied and solved by NC-TWRC with buffering in \cite{ChenONC2007} and \cite{HsuONC2011}. It was shown that the optimal transmission policy by a Markovian process formulation minimized the transmission power and symbol delay simultaneously and in the long run. In \cite{Ding2012}, the NC-TWRC in \cite{ChenONC2007,HsuONC2011} was extended to include the dynamics of wireless channels (Fig.~\ref{fig:2U1R_M_W}). In this system, a transmission policy that solves power-delay tradeoff may not be the best decision rule because it does not consider the possible loss in throughput due to the downlink transmission errors. For this reason, the scheduler is required to make optimal decision that simultaneously minimizes the transmission power, symbol delay, downlink BER in the long run by considering current queue and channel states and their expectations in the future. In \cite{Ding2012}, this problem was formulated by a discounted infinite horizon Markov decision process (MDP) with channels modeled by finite-state Markov chains (FSMCs)\cite{Sadeghi2008}. The optimal transmission policy searched by dynamic programming (DP) was shown to be superior to \cite{ChenONC2007} and \cite{HsuONC2011} in terms of enhancing the QoS (quality of service, evaluated by symbol delay and overflow in the data link layer and power consumption and error rate in the physical layer) in a practical wireless environment, e.g., Rayleigh fading channels.

However, the DP algorithm in \cite{Ding2012} is burdened with high complexity. The scheduler requires information of both queue states/occupancies and channel states to assist the decision making, i.e., the system state is a $4$-tuple (two channels and two queues) and the decision/action is a $2$-tuple (each associated with the departure control of one queue). In such a high dimensional MDP, \emph{the curse of dimensionality}\footnote{The complexity of the algorithm grows exponentially with the cardinality of the system variables \cite{SuttonRL1998}.} becomes more evident. The problem could be intractable if the cardinality of any tuple in the state variable is large. To relieve the curse, one solution is to qualitatively understand the model and prove the existence of a structured optimal policy \cite{Smith2002}. Then a modified DP algorithm, or an alternative algorithm with lower complexity, can be proposed. For example, if the optimal policy is proved to be nondecreasing in the state variable, we can run a monotonic policy iteration \cite{Djonin2007}, which reduces the computation load by shrinking the feasible action space with the increasing state index in each iteration of DP; if the optimal policy is of threshold type, the problem can be solved by a simultaneous perturbation stochastic approximation (SPSA) algorithm \cite{Huang2010}. But, structured optimal policy does not exist in general. Most often, optimal policy exists, but it varies with the state variable irregularly. In order to prove the existence of certain feature (e.g., monotonicity) in the optimal policy, we need to extensively analyze the MDP model and the recursive functions in DP algorithm. The basic approach in the existing literature is to show by induction that the monotonicity/submodularity is preserved in each iterative optimization process (maximization/minimization) in DP, e.g., \cite{Huang2010,Djonin2007}. We adopt the same method in this paper, but consider these properties in high dimensional cases. For example, a 3-dimensional submodularity (instead of 2-dimensional as usual) is shown to contribute to a monotonic optimal solution in both queue and channel states. Moreover, we use $\LN$-convexity and multimodularity, two concepts that were originally defined in discrete convex analysis \cite{Topkis2001,Murota2005} and recently applied in operational research \cite{Zhuang2012,Zipkin2008,Pang2012}, to describe the joint submodularity and integral convexity in a high dimensional space.

The aim of our work is to establish the sufficient conditions for the existence of a monotonic optimal transmission policy in the NC-TWRC system in Fig.~\ref{fig:2U1R_M_W}. Unlike other related research work where certain features were assumed \textit{a priori} (e.g., strict submodularity of DP functions \cite{Hoang2008} or uniformly distributed traffic flows \cite{Huang2010}), we prove the properties of DP by observing the variation of parameters in the immediate cost function of MDP (e.g., unit costs associated with symbol delay and transmission power, quantized error rate associated with the channel state, etc.) while having our assumptions consistent with the actual applications (e.g., arbitrary traffic rates, flat and slow Rayleigh fading channels). The main results in this paper are:
\begin{itemize}
    \item We prove that each tuple in the optimal policy is nondecreasing in the queue state that is controlled by that tuple if the chosen values of unit costs give rise to an $\LN$-convex or multimodular DP. Moreover, we show that the same results found in \cite{Huang2010} and \cite{Djonin2007} can also be explained by $\LN$-convexity or multimodularity by a unimodular coordinate transform \cite{Yu2013}.
    \item By thinking of each iteration in DP as a one-stage pure coordination supermodular game, we show that equiprobable traffic rates and certain conditions on unit costs guarantee that each tuple in the optimal policy is monotonic in not only the queue state that is controlled by that tuple but also the queue state that is associated with the information flow of the opposite direction, i.e. the one that is not under the control of that tuple.
    \item By observing the submodularity of DP, we show the sufficient conditions for an optimal policy to be nondecreasing in both queue and channel states in terms of unit costs, channel statistics, FSMC models and modulation scheme.
\end{itemize}

The rest of this paper is organized as follows. In Section~\ref{sec:sys}, we state the optimization problem in NC-TWRC with random symbol arrivals and FSMC modeled channels and clarify the assumptions of this system. In Section~\ref{sec:MDP}, we describe the MDP formulation, state the objective and present the DP algorithm. In Section~\ref{sec:Struct}, we investigate the structure in the optimal transmission policy found by DP algorithm in queue and channel states, where numerical examples are presented.

\section{System}
\label{sec:sys}

\subsection{NC-TWRC with random symbol arrivals and wireless fading channels}

Consider the NC-TWRC shown in Fig.~\ref{fig:2U1R_M_W}. User 1 and 2 randomly send symbols to each other via the relay. The relay is equipped with two finite-length FIFO queues, queue $1$ and $2$, to buffer the incoming symbols from user $1$ and $2$, respectively. The outflows of queues are controlled by a scheduler. The scheduler keeps making decisions as to whether or not to transmit symbols from queues. If the decision results in a pair of symbols in opposite directions transmitted at the same time, they will be XORed (coded) and broadcast. Otherwise, the symbol will be simply forwarded to the end user. The optimization problem is to find an optimal transmission control policy that minimizes symbol delay, queue overflow, transmission power (saved by utilizing the coding opportunities) and downlink transmission errors simultaneously and their expectations in the future.

Obviously, the optimization concerns are contradictory to each other: (1) Because of the random symbol inflows, there would not always be a pair of symbols for XORing. In this situation, waiting for coding opportunity by holding symbols results in a high symbol delay on average, while transmitting a symbol without coding results in one more symbol to be transmitted in the future, i.e., more transmission power on average; (2) Even if there is a coding opportunity, broadcasting a coded symbol when any of the channels is having low SNR will incur a high symbol error rate. Therefore, the scheduler must seek an optimal decision rule that solves this \textit{power-delay-error tradeoff}.

\subsection{Assumptions}

We consider a discrete-time decision making process, where the time is divided into small intervals, called \textit{decision epochs} and denoted by $t\in\{0,1,\dotsc,T\}$. Let $i\in\{1,2\}$ and assume the following:
\begin{enumerate}[{A}1]

\item (\IID \ incoming traffic) Denote random variable $\Fit\in{\F_i}$ as the number of incoming symbols to queue $i$ at decision epoch $t$. Let the maximum number of symbols arrived per decision epoch be no greater than $1$, i.e., \FDef. Assume that $\{f_1^{(t)}\}$ and $\{f_2^{(t)}\}$ are two independent \IID\ random processes with $Pr(\Fit=1)=p_i$ and $Pr(\Fit=0)=1-p_i$ for all $t$. Since the traffic flow rate is maximum one symbol per decision epoch, the length of decision epoch is equal to the symbol duration.

\item (finite-length queues) Queue $i$ can store maximum $L_i$ symbols. At each $t$, the scheduler makes a decision and incurs an immediate cost before the event $\mathbf{f}^{(t)}=(f_1^{(t)},f_2^{(t)})$. Denote $b_i^{(t)}\in{\B_i}$ as the occupancy of queue $i$ at the beginning of decision epoch $t$, then $\B_i=\{0,1,\dotsc,L_i+\max{\{f_i^{(t-1)}\}}\}=\{0,1,\dotsc,L_i+1\}$. If the relay's decision results in queue occupation $L_i+1$, the newly arrived symbol will be dropped. We call it symbol lost due to the \textit{queue overflow}.

\item (Markovian channel modeling) Let the full variation range of $\snr_i^{(t)}$, the instantaneous SNR of channel $i$, be partitioned into $K_i$ non-overlapping regions $\{[\SNR_1,\SNR_2),[\SNR_2,\SNR_3),\dotsc,[\SNR_{K_i},\infty)\}$, called channel states. Denote $\G_i=\{1,2,\dotsc,K_i\}$ as the state set of channel $i$ and $g_i^{(t)}$ as the state of channel $i$ at decision epoch $t$. We say that $g_i^{(t)}=k_i$ if $\snr_i^{(t)}\in{[\SNR_{k_i},\SNR_{k_i+1})}$. Each channel is modeled by a finite-state Markov chain (FSMC) \cite{Sadeghi2008}, where the state evolution of channel $i$ is governed by the transition probability $P_{g_i^{(t)}g_i^{(t+1)}}=Pr(g_i^{(t+1)}|g_i^{(t)})$.

\item (downlink channel state information) Let $\{g_1^{(t)}\}$ and $\{g_2^{(t)}\}$ be two independent and \textit{i.i.d.}\ random processes. The relay has the channel state information (channel state $g_i^{(t)}$ and transition probability $P_{g_i^{(t)}g_i^{(t+1)}}$) of both channels before the decision making at $t$.

\end{enumerate}

\section{Markov Decision Process Formulation}
\label{sec:MDP}

Based on A1, A3 and A4, we know that the statistics of the incoming traffic flow and channel dynamics associated with user $1$ or $2$ are time-invariant. It follows that the transmission control problem in Fig.~\ref{fig:2U1R_M_W} can be formulated as a stationary Markov decision process (MDP). In the following context, we drop the decision epoch notation $t$ in A1-A4 and use the notation $y$ and $y'$ for the system variable $y$ at the current and next decision epochs, respectively.

\subsection{System State}

Denote the system state $\x=(\bv, \gv)\in{\X}$, where $\bv=(b_1,b_2)\in{\B_1\times{\B_2}}$ and $\gv=(g_1,g_2)\in{\G_1\times{\G_2}}$, i.e., $\X=\B_1\times{\B_2}\times{\G_1}\times{\G_2}$. $\times$ denotes the Cartesian product. We also use the 4-tuple notation $\x=(b_1,b_2,g_1,g_2)$ in the following context.

\subsection{Action}

Denote action $\act=(a_1,a_2)\in{\A}$, where $a_i\in{\A_i}=\{0,1\}$ denotes the number of symbols departed from queue $i$ and $\A=\A_1\times{\A_2}=\{0,1\}^2$. The terminology of actions are shown in Table~\ref{tab:ActionSet}.

\begin{table}[tbp]
	\renewcommand{\arraystretch}{1.3}
	\caption{Action Set}
	\label{tab:ActionSet}
	\centering
	\begin{tabular}{c c}
	\hline\hline
	$\act$& \bfseries action itemize \\ %
	\hline
	$(0,0)$ & no transmission \\
	$(1,0)$ & forward one symbol in queue 1  \\
	$(0,1)$ & forward one symbol in queue 2 \\
	$(1,1)$ & XOR two symbols one in each queue, then broadcast.\\[0.3ex] 
	\hline 
	\end{tabular}
\end{table}

\subsection{State Transition Probabilities}

The transition probability $\Pxx=Pr(\x|\x',\act)$ denotes the probability of being in state $\x'$ at next decision epoch if action $\act$ is taken in state $\x$ at current decision epoch. Due to the assumptions of independent random processes in A1 and A4, the state transition probability is given by
\begin{equation}
    \Pxx=\Pbbv\Pggv=\prod_{i=1}^{2}\Pbbi\Pggi,
\end{equation}
where $\Pggi$ is determined by channel statistics and FSMC modeling method in A3 and $\Pbbi$ is the queue state transition probability derived as follows.

At current decision epoch, the occupancy of queue $i$ after decision $a_i$ is $\min\{[b_i-a_i]^{+},L_i\}$, where $[y]^{+}=\max\{y,0\}$. Then, the occupancy at the beginning of the next decision epoch is given by
\begin{equation}
    b'_i=\min\{[b_i-a_i]^{+},L_i\}+f_i.
\end{equation}
Therefore, the state transition probability of queue $i$ is
\begin{align}		
\Pbbi&=Pr\left(f_i=b_i^{\prime}-\min\{[b_i-a_i]^{+},L_i\}\right)                        \nonumber \\
     &=Pr\left(f_i=b_i^{\prime}-[b_i-a_i]^{+}+\IND_{\{[b_i-a_i]^{+}>L_i\}}\right)       \nonumber \\
     &=\begin{cases}		
		  Pr(f_i=b_i^{\prime}-[b_i-a_i]^{+}) & [b_i-a_i]^{+}\leq{L_i}\\
		  Pr(f_i=b_i^{\prime}-L_i) & [b_i-a_i]^{+}>L_i
	   \end{cases},
\end{align}	
where $\IND_{\{\cdot\}}$ is the indicator function that returns $1$ if the expression in $\{\cdot\}$ is true and $0$ otherwise.

\subsection{Immediate Cost}

$C:\X\times\A\rightarrow{\R_+}$ is the cost incurred immediately after action $\act$ is taken in state $\x$ at current decision epoch. It reflects three optimization concerns: the symbol delay and queue overflow, the transmission power and the downlink transmission error rate.

\subsubsection*{Holding and overflow cost}

We define $h_i$, the holding and queue overflow cost associated with queue $i$, as
  \begin{align}
        h_i(y_i)&=\lambda\min\{[y_i]^+,L_i\}+\xi_o\IND_{\{[y_i]^+=L_i+1\}}  \nonumber \\
                        &=\lambda[y_i]^{+}+(\xi_o-\lambda)\IND_{\{[y_i]^{+}=L_i+1\}}.
  \end{align}
  $\lambda>0$ is the unit holding cost and $\xi_o>\lambda$ is the unit queue overflow cost, which makes $h_i(y_i)$ a nondecreasing convex function. In the case when $y_i=b_i-a_i$, $\min\{[y_i]^+,L_i\}$ and $\IND_{\{[y_i]^+=L_i+1\}}$ count the number of symbols held in queue $i$ and the number of symbols lost due the overflow of queue $i$, respectively. We say that the term $\lambda\min\{[y_i]^+,L_i\}$ accounts for the symbol delay because by Little's Law the average symbol delay is proportional to the average number of symbols held in the queue in the long run for a given symbol arrival rate \cite{Yang2012}. We sum up $h_i$ for $i\in\{1,2\}$ and obtain the total holding and overflow cost as
  \begin{equation} \label{eq:HoldCost}
        C_{h}(\bv,\act)=\sum_{i=1}^{2}h_i(b_i-a_i).
  \end{equation}

\subsubsection*{Transmission cost}

Since forwarding and broadcasting one symbol, either coded or non-coded, consume the same amount of power, we have the immediate transmission cost as
  \begin{equation}\label{eq:TransCost}
        t_r(\act)=\tau\IND_{\{a_1=1 \text{ or }a_2=1\}}=
            \begin{cases}		
		      0 & \act=(0,0)\\
		      \tau & \text{otherwise}
	       \end{cases},
  \end{equation}
where $\tau>\lambda$ is the unit transmission cost and $\IND_{\{a_1=1 \text{ or }a_2=1\}}$ counts the number of transmissions resulting from action $\act$.

Note that \eqref{eq:HoldCost} and \eqref{eq:TransCost} form a power-delay tradeoff. A policy that always transmits whenever there is an incoming symbol without considering coding opportunities in the long run is penalized by \eqref{eq:TransCost}, and a policy that always holds symbol to wait for coding opportunities without considering the average symbol delay is penalized by \eqref{eq:HoldCost}.

\subsubsection*{Symbol error cost}

Let $P_e(g_i)$ denote the symbol error probability when channel $i$ is in state $g_i$. The form of the function $P_e$ is determined by the modulation scheme (e.g., $P_e(g_i)=\frac{1}{2}\erfc(\sqrt{\SNR_{g_i}})$ for BPSK modulation). And, $P_e(g_i)\leq{0.5}$ for all $g_i$ because $\SNR_1\geq{0}$ in A3. Since symbol errors happen only when we decide to transmit, we define the immediate symbol error cost due to the action $a_i$ as
  \begin{equation} \label{eq:ErrCost}
        \err(g_{-i},a_i)=\eta a_iP_e(g_{-i}),
  \end{equation}
where $\eta$ is the unit symbol error cost and $-i\in\{1,2\}\setminus\{i\}$, i.e., $-i=2$ if $i=1$, and $-i=1$ if $i=2$. Note, the reason we have $\err(g_{-i},a_i)$ is because the symbol departing queue $i$ is transmitted through channel $-i$, e.g., the relay sends one symbol in queue $1$ through fading channel $2$ when $a_1=1$.

Note, the aforementioned power-delay tradeoff formed by \eqref{eq:HoldCost} and \eqref{eq:TransCost} just poses the problem: whether or not to transmit if an instantaneous symbol inflow is not able to form an XORing pair. However, if the scheduler considers downlink transmission error rate in addition, a policy that always broadcasts XORed symbols whenever there's a coding opportunity without considering downlink channel states is penalized by \eqref{eq:ErrCost}. Therefore, \eqref{eq:HoldCost}, \eqref{eq:TransCost} and \eqref{eq:ErrCost} form a power-delay-error tradeoff.

In summary, we define the immediate cost as
\begin{equation}
        C(\x,\act)=C(\bv,\gv,\act)=C_{h}(\bv,\act)+C_{t}(\gv,\act),
\end{equation}
where
\begin{align}
    C_{t}(\gv,\act) &= \sum_{i=1}^{2}\err(g_{-i},a_i)+t_r(\act) \nonumber\\
    &= \eta\sum_{i=1}^{2}a_iP_e(g_{-i})+\tau\IND_{\{a_1=1\ or\ a_2=1\}}.
\end{align}

Here, $C(\x,\act)$ is in fact a linear combination of loss functions (each quantifies an optimization concern). The unit cost $\lambda$, $\xi_o$, $\tau$ and $\eta$ can be considered as the weight factors that are either given or adjustable depending on the real applications. In Section~\ref{sec:Struct}, we will derive the sufficient conditions of the existence of a structured optimal policy mainly in terms of the chosen values of these unit costs.

\subsection{Objective and Dynamic Programming}

Let $\x^{(t)}$ and $\act^{(t)}$ denote the state and action at decision epoch $t$, respectively, and consider an infinite-horizon MDP modeling where the discrete decision making process is assumed to be infinitely long. We can describe the long-run objective as
\begin{align} \label{eq:obj}
        \min\E\left[\sum_{t=0}^{\infty} \beta^t C(\x^{(t)},\act^{(t)})|\x^{(0)}\right], \forall{\x^{(0)}\in{\X}},
\end{align}
where $\beta\in[0,1)$ is the discounted factor that ensures the convergence of the series. It is proved in \cite{PutermanMDP1994}, that if the state space $\X$ is countable, the action set $\A$ is finite and the MDP is stationary, there exists an optimal deterministic stationary policy $\theta^*:\X\rightarrow{\A}$, and $\theta^*$ can be searched by dynamic programming (DP)
\begin{align}
    &Q^{(n)}(\x,\act)=C(\x,\act)+\beta\smashoperator{\sum_{\x'\in{\X}}} \Pxx  V^{(n-1)}(\x'),      \\
    &V^{(n)}(\x)=\min_{\act\in{\A}}Q^{(n)}(\x,\act),
\end{align}
where $n$ denotes the iteration index, $V^{(0)}=0$ for all $\x$ and the optimal policy $\theta^*(\x)=\arg\min_{\act\in{\A}}Q^{(N)}(\x,\act)$ if DP converges at $N$th iteration. Usually a very small convergence threshold is applied, i.e., $N<\infty$.

\section{Structured Optimal Policies}
\label{sec:Struct}

A conventional transmission control MDP model, say \cite{Hoang2008} for adaptive modulation purpose, usually has $2$-tuple state and $1$-tuple action variables. However, the MDP model defined in Section~\ref{sec:MDP} has a higher dimension: $4$-tuple state and $2$-tuple action. This could make the DP algorithm cumbersome. The major problem with DP is the \textit{curse of dimensionality}, the computation load grows exponentially with the number and dimensions of system parameters. The consequence is that the optimization problem may become intractable. For example, an increment in the cardinality of any tuple in the state variable $\x=(b_1,b_2,g_1,g_2)$ may severely overload the CPU in DP iterations or, even worse, drive the processer out of memory during MDP modeling. To cope with this problem, researchers are always interested in certain structures, e.g., monotonicity, in the optimal policy because a modified optimization algorithm with lower complexity, e.g., structured policy iteration \cite{Djonin2007} or simultaneous perturbation stochastic approximation (SPSA) \cite{Huang2010}, can be proposed if so. In this section, we investigate the submodularity, $\LN$-convexity and multimodularity of functions $Q^{(n)}(\x,\act)$ and $V^{(n)}(\x)$ in DP to establish the sufficient conditions for the existence of a monotonic optimal policy. Before that, we clarify some concepts\footnote{The definitions and lemmas related to submodularity are based on functions defined on a lattice \cite{Topkis2001}. We omitted the notation of ``lattice'' because it is obvious that queue/channel state space in our model is a complete lattice, which can be verified using the definition in \cite{Topkis1978}.} as follows.

\begin{definition}[Monotonic policy] \label{def:monotone}
Let $\theta\colon\Z^n\rightarrow{\Z^m}$, $\theta(\x)$ is monotonic nondecreasing if $\theta(\x^+)\succeq{\theta(\x^-})$, for all $\x^+,\x^-\in{\Z^n}$ such that $\x^+\succeq{\x^-}$, where $\succeq$ denotes componentwise greater than or equal to.
\end{definition}

In the case when $|\X|\gg{|\A|}$, a nondecreasing optimal policy is in the form of
    \begin{equation}
        \theta^{*}(\x)= \begin{cases}
                            \act_{|\A|}  &  \phi_{\act_{|\A|}}^*\preceq{\x}   \\
                            \qquad \vdots   \\
                            \act_2       &  \phi_{\act_1}^*\preceq{\x}\prec{\phi_{\act_2}^*}  \\
                            \act_1       &  \x\prec{\phi_{\act_1}^*}
                         \end{cases}.
    \end{equation}
$\theta^{*}$ is a switching curve or plane that is characterized by optimal threshold vector
\begin{equation}
    \Thv=(\phi_{\act_1}^*,\phi_{\act_2}^*,\cdots,\phi_{\act_{|\A|}}^*).
\end{equation}
In this case, we can formulate a multivariate optimization problem in the form of
\begin{equation}
    \min_{\Thv}J(\Thv)
\end{equation}
with $J$ being the objective in \eqref{eq:obj}. This problem can be solved by an SPSA algorithm, e.g., \cite{Huang2010}.

\begin{definition}[Submodularity] \label{def:submodularity}
Let $\ev_i\in{\Z^n}$ be an $n$-tuple with all zero entries except the $i$th entry being one. $f\colon\Z^n\rightarrow{\R_+}$ is submodular if $f(\x+\ev_i)+f(\x+\ev_j)\geq{f(\x)+f(\x+\ev_i+\ev_j)}$ for all $\x\in{\Z^n}$ and $1 \leq i,j \leq n$ such that $i\neq{j}$. $f$ is strictly submodular if the inequality is strict.\footnote{This definition is based on the submodularity defined on lattice in \cite{Murota2005} and Proposition 2.1 in \cite{Hajek1985}}
\end{definition}

The insight of the submodularity can be explained by the following example. In DP, a submodular function $Q^{(n)}(\x,\act)$ has $Q^{(n)}(\x,\act^-)-Q^{(n)}(\x,\act^+)$ nondecreasing in $\x$ for all $\act^+\succeq{\act^-}$\footnote{According to Definition~\ref{def:submodularity}, $Q(\x,\act)$ is submodular if $Q(\x^+,\act^-)-Q(\x^+,\act^+)\geq{Q(\x^-,\act^-)-Q(\x^-,\act^+)}$ for all $\act^+\succeq{\act^-}$ and $\x^+\succeq{\x^-}$.}, i.e., the preference of choosing action $\act^+$ over $\act^-$ is always nondecreasing in $\x$. Therefore, an increase in the state variable $\x$ implies an increase in the decision rule $\theta^{(n)}(\x)=\min_{\act}Q^{(n)}(\x,\act)$. It follows that the optimal policy $\theta^*(\x)=\min_{\act}Q^{(N)}(\x,\act)$ must be monotonic in $\x$ if we can prove the submodularity of $Q^{(n)}$ for all $n$. This property is summarized in a general form in the following lemma.

\begin{lemma}  \label{lemma:SubmodularPreserve}
Submodularity has the following properties:
\begin{enumerate}[(a)]
  \item If $g\colon\Z^n\rightarrow{\R_+}$ is submodular in $(\x,\y)\in{\Z^n}$, then $f(\x)=\min_{\y}g(\x,\y)$ is submodular in $\x$, and the minimizer $\y^*(\x)=\arg\min_{\y}g(\x,\y)$ is nondecreasing in $\x$. (Theorem 4.3 and 6.1 in \cite{Topkis1978})
  \item If $f_i\colon\Z^n\rightarrow{\R_+}$ is submodular in $\x\in{\Z^n}$ and $\alpha_i\geq{0}$ for all $i$, then $\sum_{i=1}^{m} \alpha_i f_i(\x)$ is submodular in $\x$. (Proof in Appendix~\ref{app:SubmodularPreserve}) \hfill\IEEEQED
\end{enumerate}
\end{lemma}

\begin{definition}[$\LN$-convexity \cite{Murota2005}] \label{def:Lconvex}
$f\colon\Z^n\rightarrow{\R_+}$ is $\LN$-convex if $\psi(\x,\zeta)=f(\x-\zeta\One)$ is submodular in $(\x,\zeta)$, where $\One=(1,1,\dotsc,1)\in{\Z^n}$ and $\zeta\in{\Z}$.
\end{definition}

\begin{definition}[multimodularity \cite{Murota2005}] \label{def:multimodular}
$f\colon\Z^n\rightarrow{\R_+}$ is multimodular if $\psi(\x,\zeta)=f(x_1-\zeta,x_2-x_1,\dotsc,x_n-x_{n-1})$ is submodular in $(\x,\zeta)$, where $\zeta\in{\Z}$.
\end{definition}

$\LN$-convexity and multimodularity are two concepts defined in discrete convex analysis \cite{Murota2003}. $\LN$-convexity and multimodularity both imply integral convexity. The difference is that $\LN$-convexity is submodular while multimodularity is supermoduar\footnote{$f\colon\Z^n\rightarrow{\R_-}$ is (strictly) supermodular if $-f$ is (strictly) submodular.} \cite{Yu2013}. The two concepts are related by a unimodular coordinate transform.

\begin{lemma}[unimodular coordinate transform \cite{Yu2013}] \label{lemma:UniTrans}
Let matrix $M_{n,i}=\left[ \begin{array}{ccc}
-U_i & 0  \\
0 & L_{n-i}  \end{array} \right]$, where $U_i$ and $L_{i}$ are the $i\times{i}$ upper and lower triangular matrix with all nonzero entries being one, respectively, then
\begin{enumerate}[(a)]
  \item a function $f\colon\Z^n\rightarrow{\R_+}$ is multimodular if and only if it can be represented by $f(\x)=g(\pm M_{n,i}\x)$ for some $\LN$-convex function $g$.
  \item a function $g\colon\Z^n\rightarrow{\R_+}$ is $\LN$-convex if and only if it can be represented by $g(\x)=f(\pm M_{n,i}^{-1}\x)$ for some multimodular function $f$. \hfill\IEEEQED
\end{enumerate}
\end{lemma}

$\LN$-convexity and multimodularity are properties commonly seen in MDP modeled flow control problems in queueing networks \cite{Altman2000} and inventory systems \cite{Zipkin2008}. Due to the implication of the submodularity/supermodularity, they both contribute to a monotonic structure in the optimal policy. In section~\ref{sec:StructB}, we will use them to show a nondecreasing optimal transmission policy in queue states. As a preliminary step, we clarify some properties of $\LN$-convexity and multimodularity in the following lemma.

\begin{lemma}  \label{lemma:LconvexPreserve}
$\LN$-convexity and multimodularity have the following properties:
\begin{enumerate}[(a)]
  \item If $g\colon\Z^n\rightarrow{\R_+}$ is $\LN$-convex/multimodular in $(\x,\y)\in\Z^n$, then $f(\x)=\min_{\y}g(\x,\y)$ is $\LN$-convex/multimodular in $\x$, and the minimizer $\y^*(\x)=\arg\min_{\y}g(\x,\y)$ is nondecreasing/nonincreasing in $\x$. (Proof in Appendix~\ref{app:SubmodularPreserve})
  \item If $f_i\colon\Z^n\rightarrow\R_+$ is $\LN$-convex/multimodular and $\alpha_i\geq{0}$ for all $i$, then $\sum_{i=1}^{m} \alpha_i f_i(\x)$ is $\LN$-convex/multimodular in $\x$. (Proof in Appendix~\ref{app:SubmodularPreserve})
  \item If $h\colon\Z\rightarrow{\R_+}$ is convex\footnote{The one dimensional discrete convex function $h:\Z\rightarrow{\R}$ satisfies $h(x+1)+h(x-1)-2h(x)\geq{0}$ for all $x\in{\Z}$. Moreover, by Definition~\ref{def:Lconvex} and \ref{def:multimodular}, $h$ is both $\LN$-convex and multimodular.}, then $f(\x)=h(x_1-x_2)$ is $\LN$-convex in $\x=(x_1,x_2)$ and $g(\x)=h(x_1+x_2)$ is multimodular in $\x=(x_1,x_2)$. (Proof in Appendix~\ref{app:SubmodularPreserve})
  \item Let $d$ be a random variable. If $g(\x,d)$ is $\LN$-convex/multimodular in $\x\in{\Z^n}$ for all $d$, then $\E_{d}[g(\x,d)]$ is $\LN$-convex/multimodular in $\x$. \cite{Zipkin2008,Yu2013}
  \item If $f\colon\Z^n\rightarrow\R_+$ is $\LN$-convex, then $\psi(\x,\zeta)=f(\x-\zeta\One)$ is $\LN$-convex in $(\x,\zeta)$. (Lemma 1 in \cite{Zipkin2008})  \hfill\IEEEQED
\end{enumerate}
\end{lemma}

\begin{definition}[First order stochastic dominance \cite{Smith2002}] \label{def:1stStoDom}
Let $\tilde{\rho}(x)$ be a random selection on space $\X$ according to a probability measure $\mu(x)$ where $x$ conditions the random selection, then $\tilde{\rho}(x)$ is first order stochastically nondecreasing in $x$ if $\E[u(\tilde{\rho}(x^+))] \geq \E[u(\tilde{\rho}(x^-))]$ for all nondecreasing functions $u$ and $x^+ \geq x^-$.
\end{definition}

Stochastic dominance, as defined in decision theory, describes the situation that the expected aftermath (quantified by a utility or cost function) of one decision is superior to that of another. In Section~\ref{sec:StructG}, we will show that the stochastic dominance of the channel state transition probabilities preserves submodularity across the iterations in DP and contributes to a nondecreasing optimal transmission policy in channel states.

\subsection{Structured Properties of Dynamic Programming}

To propose the prototypical procedure of proving the existence of a monotonic optimal policy, we first define a $\Pro$ property as follows:
\begin{definition}[$\Pro$ property] \label{def:PstarProperty}
$f\colon\Z^n\rightarrow{\R_+}$ has $\Pro$ property in $(\x,\y)\in{\Z^n}$ if $f^*(\x)=\min_{\y}f(\x,\y)$ has $\Pro$ property in $\x$ and $\y^*(\x)=\arg\min_{\x}f(\x,\y)$ is monotonic (nondecreasing/nonincreasing) in $\x$.
\end{definition}

It can be seen, by Lemma~\ref{lemma:SubmodularPreserve}(a) and \ref{lemma:LconvexPreserve}(a), that submodularity, $\LN$-convexity and multimodularity satisfy the conditions in Definition~\ref{def:PstarProperty}, which we summarize as follows.
\begin{theorem} \label{theo:PstarProperty}
Submodularity, $\LN$-convexity and multimodularity have $\Pro$ property.  \hfill\IEEEQED
\end{theorem}

We therefore propose an approach, similar to Proposition 5 in \cite{Smith2002}, as follows:
\begin{proposition} \label{prop:PropertyDP}
Let DP converge at $N$th iteration. The optimal value function $V^{*}(\x)=V^{(N)}(\x)$ has $\Pro$ property, and the optimal policy $\theta^*$ is monotonic in $\x$, if:
\begin{itemize}
  \item[(a)] $C(\x,\act)$ has $\Pro$ property,
  \item[(b)] $Q^{(n)}(\x,\act)=C(\x,\act)+\beta\sum_{\x'\in{\X}}\Pxx V^{(n-1)}(\x')$ has $\Pro$ property for all $\Pro$ property functions $V^{(n-1)}$ and $n$.
\end{itemize}
\end{proposition}
\begin{IEEEproof}
Since DP starts from $V^{(0)}(\x)=0$ for all $\x\in{\X}$, $Q^{(1)}=C(\x,\act)$ has $\Pro$ property. So $V^{(1)}(\x)=\min_{\act\in{\A}}Q^{(1)}(\x,\act)$ has $\Pro$ property. By induction, assume $V^{(n-1)}(\x,\act)$ has $\Pro$ property. Then $Q^{(n)}$ and $V^{(n)}(\x)=\min_{\act\in{\A}}Q^{(n)}(\x,\act)$ have $\Pro$ property. Therefore, $Q^{(N)}(\x,\act)$ and $V^{*}(\x)=V^{(N)}(\x)$ must also possess $\Pro$ property, and $\theta^*(\x)=\arg\min_{\act\in{\A}}Q^{(N)}(\x,\act)$ is monotonic in $\x$.
\end{IEEEproof}

\subsection{Monotonic Policies in Queues States}
\label{sec:StructB}

\subsubsection{Nondecreasing $a_i^*$ in $b_i$}

Let the optimal action be $\act^*=(a_1^*,a_2^*)=\theta^*(\x)$. The following theorem shows that the optimal action $a_i^*$ is monotonic in $b_i$, the state of queue being controlled by $a_i$ if the unit costs satisfy a certain condition.
\begin{theorem} \label{theo:LconvexOfImmediate}
If $\xi_o \geq{2\lambda+\eta+\tau}$, then for all $i\in\{1,2\}$ $C(\x,\act)$ and $Q^{(n)}(\x,\act)$ are nondecreasing in $b_i$ and $\LN$-convex in $(b_i,a_i)$, $V^*(\x)$ is nondecreasing and $\LN$-convex in $b_i$, and the optimal action $a^*_i$ is nondecreasing in $b_i$.
\end{theorem}
\begin{IEEEproof}
We define two functions
\begin{equation}
\tilde{C}(\y,\gv,\act)=\tilde{C}_h(\y)+C_t(\gv,\act),
\end{equation}
where $\tilde{C}_h(\y)=\sum_{i=1}^{2}h_i(y_i)$ and
\begin{equation}
\tilde{Q}^{(n)}(\y,\gv,\act)=\tilde{C}(\y,\gv,\act)+\beta \E_{\gv'} \Big[ V_\fv^{(n-1)}(\y,\gv') \Big|\gv \Big].
\end{equation}
Here,
\begin{equation}
	V_\fv^{(n-1)}(\y,\gv')=\E_\fv \Big[ V^{(n-1)}(\min\{[y_1]^+,L_1\}+f_1,\min\{[y_2]^+,L_2\}+f_2,\gv') \Big],
\end{equation}
$\y=(y_1,y_2)$ and $\fv=(f_1,f_2)$. It is easy to see that $C(\bv,\gv,\act)=\tilde{C}(\bv-\act,\gv,\act)$ and $Q^{(n)}(\bv,\gv,\act)=\tilde{Q}^{(n)}(\bv-\act,\gv,\act)$.
Since
\begin{equation}
    \left[ \begin{array}{ccc} b_i-a_i \\ a_i  \end{array} \right]=\left[ \begin{array}{ccc} 1 & -1  \\ 0 & 1  \end{array} \right] \left[ \begin{array}{ccc} b_i \\ a_i  \end{array} \right]=-M_{2,2}^{-1} \left[ \begin{array}{ccc} b_i \\ a_i  \end{array} \right],
\end{equation}
according to Lemma~\ref{lemma:UniTrans}(b), it follows that proving the $\LN$-convexity of $C(\bv,\gv,\act)$ and $Q^{(n)}(\bv,\gv,\act)$ in $(b_i,a_i)$ is equivalent to showing the multimodularity of $\tilde{C}(\y,\gv,\act)$ and $\tilde{Q}^{(n)}(\y,\gv,\act)$ in $(y_i,a_i)$. It is also clear that the monotonicity of $C(\bv,\gv,\act)$ and $Q^{(n)}(\bv,\gv,\act)$ in $b_i$ is equivalent to the monotonicity of $\tilde{C}(\y,\gv,\act)$ and $\tilde{Q}^{(n)}(\y,\gv,\act)$ in $y_i$. See Appendix~\ref{app:LconvexOfImmediate} for the proof of the monotonicity and multimodularity of $\tilde{C}(\y,\gv,\act)$ and $\tilde{Q}^{(n)}(\y,\gv,\act)$ in $y_i$ and $(y_i,a_i)$, respectively.

According to Proposition 4.7.3 in \cite{PutermanMDP1994}, $V^*(\x)$ is nondecreasing in $b_i$. By Theorem~\ref{theo:PstarProperty} and Proposition~\ref{prop:PropertyDP}, $V^*(\x)$ is $\LN$-convex in $b_i$, and $a_i^*$ is nondecreasing in $b_i$.
\end{IEEEproof}

Note, Theorem~\ref{theo:LconvexOfImmediate} aligns with the existing results in the literature, e.g., the adaptive MIMO transmission control \cite{Djonin2007} and the Markov game modeled adaptive modulation of cognitive radio \cite{Huang2010}. In fact, both of them can be explained by $\LN$-convexity. In \cite{Djonin2007}, the monotonicity of $a_i^*$ in $b_i$ was shown by the multimodularity in $(b_i,-a_i)$. But
\begin{equation}
    \left[ \begin{array}{ccc} b_i \\ -a_i  \end{array} \right]=\left[ \begin{array}{ccc} 1 & 0 \\ 0 & -1  \end{array} \right] \left[ \begin{array}{ccc} b_i \\ a_i  \end{array} \right]=-M_{2,1}^{-1}\left[ \begin{array}{ccc} b_i \\ a_i  \end{array} \right]
\end{equation}
By Lemma~\ref{lemma:UniTrans}(b), we know that if the a function is multimodular in $(b_i,-a_i)$, then it must be $\LN$-convex in $(b_i,a_i)$. Consequently, $V^{(n)}(\x)$ is integer convex in $b_i$ because $\LN$-convexity in one dimension is exactly integer convexity\footnote{In \cite{Djonin2007}, integer convexity was used to denote the one dimensional discrete convexity as explained in Lemma~\ref{lemma:LconvexPreserve}(c).}. In \cite{Huang2010}, the monotonicity of $a_i^*$ was shown by the submodularity of $Q^{(n)}$ in $(b_i,a_i)$. But, $Q^{(n)}$ is a function of $b_i-a_i$. According to Definition~\ref{def:Lconvex}, the $\LN$-convexity of $g(x_1,x_2)=f(x_1-x_2)$ in $(x_1,x_2)$ is equivalent to the submodularity of $g(x_1,x_2)$ in $(x_1,x_2)$. So $Q^{(n)}$ is also $\LN$-convex in $(b_i,a_i)$.
\subsubsection{Nondecreasing $a_i^*$ in $(b_1,b_2)$}

\begin{figure}[tbp]
	\centering
		\centerline{\scalebox{1}{\begin{tikzpicture}

\draw (-3,0.8) rectangle (3,-0.8);
\draw (-3,0) -- (3,0);
\draw (0,0.8) -- (0,-0.8);

\node at (-1.5,0.4) {$-Q^{(n)}\big(\mathbf{x},(0,0)\big)$};
\node at (1.5,0.4) {$-Q^{(n)}\big(\mathbf{x},(1,0)\big)$};
\node at (-1.5,-0.4) {$-Q^{(n)}\big(\mathbf{x},(0,1)\big)$};
\node at (1.5,-0.4) {$-Q^{(n)}\big(\mathbf{x},(1,1)\big)$};

\node at (-1.5,1.2) {$a_1=0$};
\node at (1.5,1.2) {$a_1=1$};
\node at (-3.7,0.5) {$a_2=0$};
\node at (-3.7,-0.5) {$a_2=1$};

\end{tikzpicture}}}
	\caption{Utility matrix of one-stage pure coordination game in the $n$th iteration in DP. $-Q^{(n)}:\A_1\times{\A_2}\rightarrow{\R_{-}}$ is considered the utility function for a fixed $\x$.}
	\label{fig:SupGame}
\end{figure}
	
We formulate the optimization problem in the $n$th iteration of DP by a 2-player 2-strategy game, which is called one-stage game in Fig.~\ref{fig:SupGame}. Assume that action $a_1$ is taken by player 1, and $a_2$ is taken by player 2. Obviously, it is a pure coordination game where the utility $-Q^{(n)}(\x,(a_1,a_2))$ is the same to player 1 and 2.

We prove, in Appendix~\ref{app:TeamGame}, that Fig.~\ref{fig:SupGame} is a supermodular game with utility function $-Q^{(n)}(\x,\act)$ strictly supermodular in $\act=(a_1,a_2)$ for all $\x$ and $V^{(n-1)}(\x')$ that is $\LN$-convex in $\bv'=(b'_1,b'_2)$. It is proved in \cite{Milgrom1990} that there exists at least one equilibrium $(a_1^*,a_2^*)$ in the form of pure strategy in a supermodular game. Then, we have the following theorem for the monotonicity of the optimal action $a_i^*$ in $\bv=(b_1,b_2)$.
\begin{theorem} \label{theo:LconvexOfQ}
If
\begin{enumerate}[(a)]
    \item $\xi_o \geq 2\lambda+\eta+\tau$,
    \item one-stage game (in Fig.~\ref{fig:SupGame}) has two pure strategy equilibria $(0,0)$ and $(1,1)$  for all $\x=(b_1,b_2,g_1,g_2)$ such that $b_i<L_i+1$ for all $i\in\{1,2\}$,
\end{enumerate}
then $C(\x,\act)$ and $Q^{(n)}(\x,\act)$ are $\LN$-convex in $(\bv,\act)=(b_1,b_2,a_1,a_2)$, the optimal value function $V^*(\x)$ is $\LN$-convex in $\bv=(b_1,b_2)$ and the optimal action $\act^*=(a_1^*,a_2^*)$ is nondecreasing in $\bv=(b_1,b_2)$.
\end{theorem}
The proof is in Appendix~\ref{app:LconvexOfQ}. \hfill\IEEEQED

Here is a corollary of Theorem~\ref{theo:LconvexOfQ}.
\begin{corollary} \label{lemma:BREquiProb}
If
\begin{enumerate}[(a)]
  \item $\xi_o \geq 2\lambda+\eta+\tau$,
  \item $p_1=p_2=0.5$,
  \item $\beta \leq \frac{2(\tau-\lambda)}{\tau+\eta}$,
\end{enumerate}
then Theorem~\ref{theo:LconvexOfQ} holds.
\end{corollary}
The proof is in Appendix~\ref{app:BREquiProb}. \hfill\IEEEQED

\begin{figure}[t]
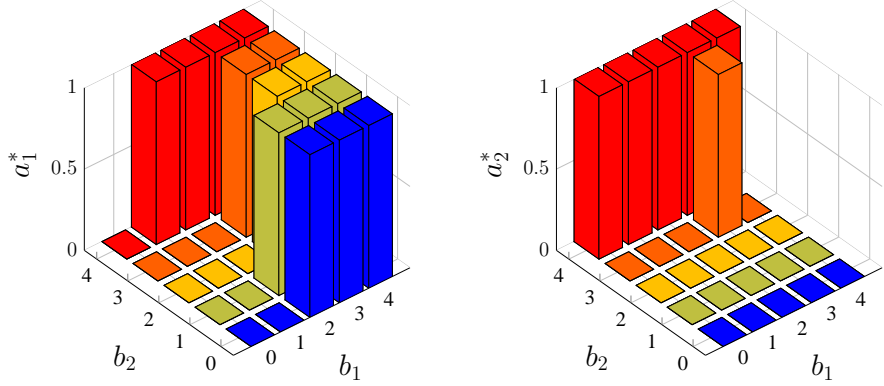

	\centering
    \subfigure{\scalebox{0.65}{\input{figures/a_b_NonMono2_1.tex}}}\qquad
    \subfigure{\scalebox{0.65}{\input{figures/a_b_NonMono2_2.tex}}}
	\caption{The optimal action $a_1^*$ (left) and $a_2^*$ (right) vs. $b_1$, the state of queue 1, and $b_2$, the state of queue 2, where $p_1=0.1$, $p_2=0.2$, $\lambda=0.05$, $\tau=1$, $\eta=2$, $\xi_o=4$ and $\beta=0.97$. In this case, $\xi_o \geq 2\lambda+\eta+\tau$. The condition in Theorem~\ref{theo:LconvexOfImmediate} is satisfied. Therefore, $a_1^*$ and $a_2^*$ are nondecreasing in $b_1$ and $b_2$, respectively.}
	\label{fig:a_b(Mono)1}
\end{figure}

\begin{figure}[tbp]
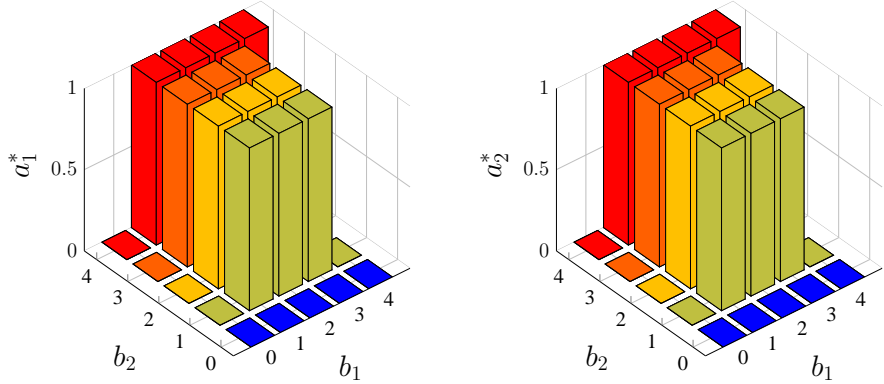

	\centering
    \subfigure{\scalebox{0.65}{\input{figures/a_b_NonMono1_1.tex}}}\qquad
    \subfigure{\scalebox{0.65}{\input{figures/a_b_NonMono1_2.tex}}}
	\caption{The optimal action $a_1^*$ (left) and $a_2^*$ (right) vs. $b_1$, the state of queue 1, and $b_2$, the state of queue 2, where $p_1=0.1$, $p_2=0.2$, $\lambda=0.05$, $\tau=1$, $\eta=2$, $\xi_o=1$ and $\beta=0.97$. In this case, $\xi_o < 2\lambda+\eta+\tau$. Theorem~\ref{theo:LconvexOfImmediate} no longer holds. As can be seen, $a_1^*$ is not monotonic in $b_1$.}
	\label{fig:a_b(Mono)2}
\end{figure}

\begin{figure}[t]
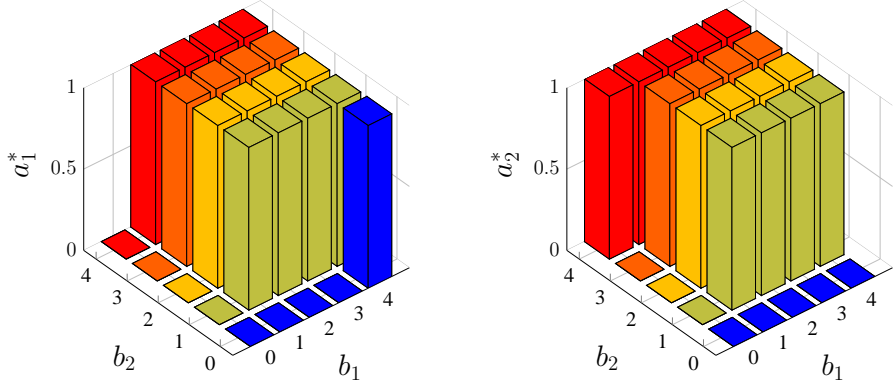

	\centering
    \subfigure{\scalebox{0.65}{\input{figures/a_bv_equiPMono_1.tex}}} \qquad
    \subfigure{\scalebox{0.65}{\input{figures/a_bv_equiPMono_2.tex}}}
	\caption{The optimal action $a_1^*$ (left) and $a_2^*$ (right) vs. $b_1$, the state of queue 1, and $b_2$, the state of queue 2, where $p_1=p_2=0.5$, $\lambda=0.05$, $\tau=1$, $\eta=1$, $\xi_o=4$ and $\beta=0.97$. In this case, $\xi_o \geq 2\lambda+\eta+\tau$ and $\beta \leq \frac{2(\tau-\lambda)}{\tau+\eta}$. According to Corollary~\ref{lemma:BREquiProb}, Theorem~\ref{theo:LconvexOfQ} holds. Therefore, both $a_1^*$ and $a_2^*$ are nondecreasing in $(b_1,b_2)$.}
	\label{fig:a_bv(Mono)1}
\end{figure}

\begin{figure}[tbp]
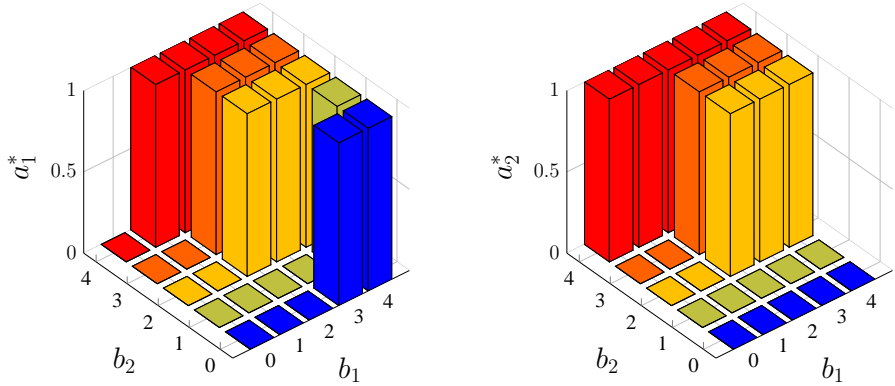

	\centering
    \subfigure{\scalebox{0.65}{\input{figures/a_bv_equiPNonMono_1.tex}}} \qquad
    \subfigure{\scalebox{0.65}{\input{figures/a_bv_equiPNonMono_2.tex}}}
	\caption{The optimal action $a_1^*$ (left) and $a_2^*$ (right) vs. $b_1$, the state of queue 1, and $b_2$, the state of queue 2, where $p_1=p_2=0.5$, $\lambda=0.05$, $\tau=1$, $\eta=2$, $\xi_o=4$ and $\beta=0.97$. In this case, $\xi_o \geq 2\lambda+\eta+\tau$ but $\beta > \frac{2(\tau-\lambda)}{\tau+\eta}$. Theorem~\ref{theo:LconvexOfImmediate} holds, while Theorem~\ref{theo:LconvexOfQ} does not. As can be seen,  $a_1^*$ and $a_2^*$ are monotonic in $b_1$ and $b_2$ respectively, but $a_1^*$ is not monotonic in $b_2$.}
	\label{fig:a_bv(Mono)2}
\end{figure}

We show examples of Theorems~\ref{theo:LconvexOfImmediate} and \ref{theo:LconvexOfQ} in Figs.~\ref{fig:a_b(Mono)1}-\ref{fig:a_bv(Mono)2}. The results are collected by value iteration, a DP algorithm, applied on an NC-TWRC system with Bernoulli symbol arrivals, $5$ queue states and $8$ channel states, i.e., $f_i^{(t)}\sim{\text{Bernoulli}(p_i)}$, $L_i=3$ and $K_i=8$ for all $t$ and $i\in\{1,2\}$. In Fig.~\ref{fig:a_b(Mono)1}, we choose the values of unit costs to make Theorem~\ref{theo:LconvexOfImmediate} hold. As shown in the figure, the optimal action $a_1^*$ and $a_2^*$ are monotonic in $b_1$ and $b_2$, respectively, i.e., $a_i^*$ is nondecreasing in the queue state that is being controlled by $a_i$. In Fig.~\ref{fig:a_b(Mono)2}, we change the value of unit cost $\xi_o$ to breach the condition in Theorem~\ref{theo:LconvexOfImmediate} so that the monotonicity of $a_i^*$ in $b_i$ is not guaranteed. In this case, $a_1^*$ that is not monotonic in $b_1$.

In Fig.~\ref{fig:a_bv(Mono)1}, we choose the equiprobable symbol arrival rates $p_1=p_2=0.5$ and the unit costs according to Corollary~\ref{lemma:BREquiProb} to make Theorem~\ref{theo:LconvexOfQ} hold. As shown in the figure, the optimal action $a_1^*$ and $a_2^*$ are both nondecreasing in $(b_1,b_2)$. As compared to Fig.~\ref{fig:a_b(Mono)1}, in this case, $a_i^*$ is also monotonic in $b_{-i}$, the queue state that is affected by the message flow and transmission control in the opposite direction, i.e., the queue state that is not controlled by $a_i$. In Fig.~\ref{fig:a_bv(Mono)2}, we switch unit cost $\eta$ from $1$ to $2$ so that Theorem~\ref{theo:LconvexOfQ} no longer holds. In this case, neither $a_1^*$ nor $a_2^*$ is monotonic in $(b_1,b_2)$. But, the condition in Theorem~\ref{theo:LconvexOfImmediate} is satisfied. Therefore, $a_1^*$ and $a_2^*$ are still nondecreasing in $b_1$ and $b_2$, respectively.

\subsection{Monotonic Policies in Channel States}
\label{sec:StructG}

The related research work in the existing literature considers the structure of the optimal policy in queue state only, e.g., \cite{Huang2010},\cite{Djonin2007} and \cite{Yang2012}. This section breaks this limitation in that we extend the investigation of the monotonicity to the channel states. The main results are summarized as follows.
\begin{theorem} \label{theo:SubOnChannel}
If
\begin{itemize}
  \item[(a)] $\xi_o \geq 2\lambda+\eta+\tau$,
  \item[(b)] $P_e(g_i)\geq{P_e(g_i+1)}$,
  \item[(c)] $\Pggi$ is first order stochastic nondecreasing in $g_i$,
  \item[(d)] $\beta\leq\frac{P_e(g_i)-P_e(g_i+1)}{\sum_{g'_i}P_{g_ig'_i} ( P_e(g'_i)-P_e(g'_i+1) ) }$.
\end{itemize}
then $C(\x,\act)$ and $Q^{(n)}(\x,\act)$ is submodular in $(b_{i},g_{-i},a_{i})$, $V^{*}(\x)$ is submodular in $(b_{i},g_{-i})$ and the optimal action $a_i^*$ is nondecreasing in $(b_{i},g_{-i})$.
\end{theorem}
The proof is in Appendix~\ref{app:SubOnChannel}.  \hfill\IEEEQED

In Theorem~\ref{theo:SubOnChannel}, conditions (b)--(d) depend on the fading statistics, the FSMC modeling method and the modulation scheme. In fact, condition (b) and (c) are not hard to satisfy if we adopt the equiprobable partition methods in the FSMCs.
\begin{corollary} \label{lemma:EquiFSMC}
Under assumption A3, if the FSMC of channel $i$ adopts equiprobable partitioning (of the full range of SNR), and channel $i$ experiences slow and flat Rayleigh fading, then condition (b) and (c) in Theorem~\ref{theo:SubOnChannel} are satisfied.
\end{corollary}
The proof is in Appendix~\ref{app:EquiFSMC}.   \hfill\IEEEQED

Finally, condition (d) depends on the modulation scheme, which should be determined in the real applications.

\begin{figure}[t]
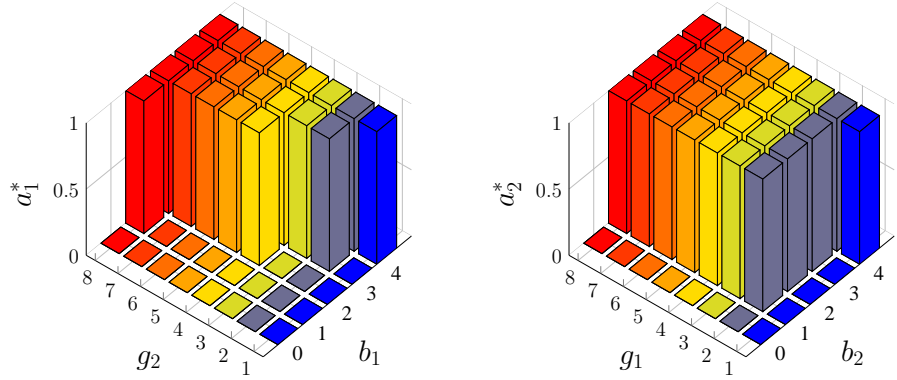

	\centering
    \subfigure{\scalebox{0.65}{\input{figures/a_bg_Mono_1.tex}}} \qquad
    \subfigure{\scalebox{0.65}{\input{figures/a_bg_Mono_2.tex}}}
	\caption{The optimal action $a_1^*$ vs. queue state $b_1$ and channel state $g_2$ (left), and $a_2^*$ vs. $b_2$ and $g_1$ (right), where $p_1=0.1$, $p_2=0.2$, $\lambda=0.05$, $\tau=1$, $\eta=2$, $\xi_o=4$ and $\beta=0.95$. Two channels are both Rayleigh fading with $\AVESNR_1=\AVESNR_2=0\si{\decibel}$ and are both modeled by 8-state equiprobable FSMCs. Modulation scheme is BPSK. In this case, $\beta\leq\frac{P_e(g_i)-P_e(g_i+1)}{\sum_{g'_i}P_{g_ig'_i} ( P_e(g'_i)-P_e(g'_i+1) ) }$, and according to Corollary~\ref{lemma:EquiFSMC}, Theorem~\ref{theo:SubOnChannel} holds. Therefore, $a_1^*$ and $a_2^*$ are nondecreasing in $(b_1,g_2)$ and $(b_2,g_1)$, respectively.}
	\label{fig:a_bg(Mono)1}
\end{figure}

\begin{figure}[tbp]
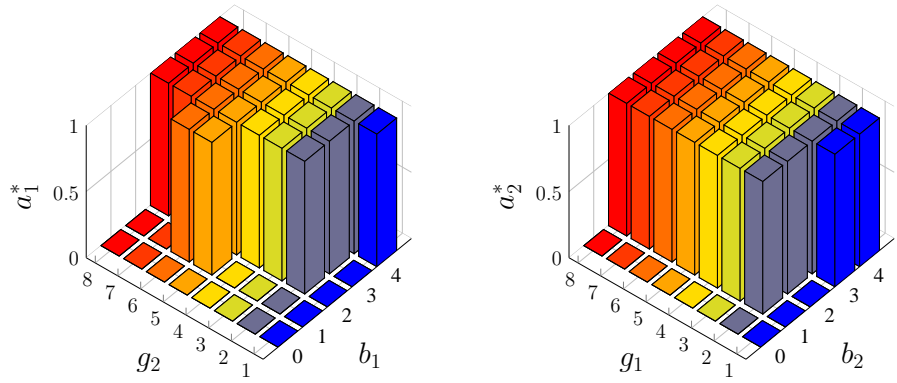

	\centering
    \subfigure{\scalebox{0.65}{\input{figures/a_bg_NonMono_1.tex}}} \qquad
    \subfigure{\scalebox{0.65}{\input{figures/a_bg_NonMono_2.tex}}}
	\caption{The optimal action $a_1^*$ vs. queue state $b_1$ and channel state $g_2$ (left), and $a_2^*$ vs. $b_2$ and $g_1$ (right), where $p_1=0.1$, $p_2=0.2$, $\lambda=0.05$, $\tau=1$, $\eta=2$, $\xi_o=4$ and $\beta=0.95$. Two channels are both Rayleigh fading and are both modeled by 8-state equiprobable FSMCs. But $\AVESNR_1=0\si{\decibel}$ and $\AVESNR_2=3\si{\decibel}$. Modulation scheme is still BPSK. In this case, $\beta\leq\frac{P_e(g_i)-P_e(g_i+1)}{\sum_{g'_i}P_{g_ig'_i} ( P_e(g'_i)-P_e(g'_i+1) ) }$ does not hold for all $g_i$. We can see that $a_1^*$ is not monotonic in $g_2$.}
	\label{fig:a_bg(Mono)2}
\end{figure}

We show examples of Theorem~\ref{theo:SubOnChannel} in Figs.~\ref{fig:a_bg(Mono)1} and \ref{fig:a_bg(Mono)2}. In Fig.~\ref{fig:a_bg(Mono)1}, we use the same system parameters as in Fig.~\ref{fig:a_b(Mono)1} except that the discount factor $\beta$ is switched from $0.97$ to $0.95$ in order to satisfy the inequality in condition (d) of Theorem~\ref{theo:SubOnChannel} . The results are obtained from an NC-TWRC system exchanging BPSK symbols over slow and flat Rayleigh fading channels with average SNR $\AVESNR_1=\AVESNR_2=0\si{\decibel}$. Both FSMCs are $8$-state and adopt equiprobable partition method. In this case, all the conditions in Theorem~\ref{theo:SubOnChannel} are satisfied according to Corollary~\ref{lemma:EquiFSMC}. Therefore, $a_1^*$ is nondecreasing in $(b_1,g_2)$, and $a_2^*$ is nondecreasing in $(b_2,g_1)$. In Fig.~\ref{fig:a_bg(Mono)2}, we switch $\AVESNR_2$ from $0\si{\decibel}$ to $3\si{\decibel}$ to breach condition (d) in Theorem~\ref{theo:SubOnChannel}. In this case, $a_1^*$ is not monotonic in $g_2$. But, since Theorem~\ref{theo:LconvexOfImmediate} still holds, $a_1^*$ and $a_2^*$ are monotonic in $b_1$ and $b_2$, respectively.

\begin{figure}[t]
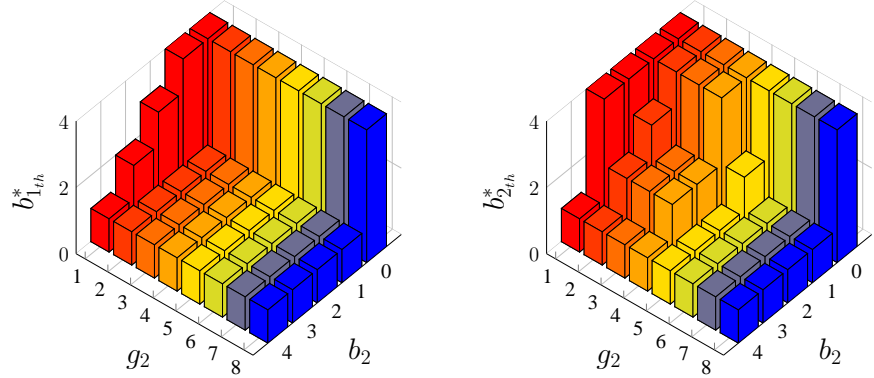

	\centering
    \subfigure{\scalebox{0.65}{\input{figures/TH_bg_Mono_1.tex}}}  \qquad
    \subfigure{\scalebox{0.65}{\input{figures/TH_bg_Mono_2.tex}}}
	\caption{The optimal threshold $b_{1_{th}}^{*}$ vs. queue state $b_{2}$ and channel state $g_{2}$ (left), and $b_{2_{th}}^{*}$ vs. $b_{1}$ and $g_{1}$ (right), with $p_1=p_2=0.5$, $\lambda=0.05$, $\tau=1$, $\eta=1$, $\xi_o=4$ and $\beta=0.95$. Two channels are both Rayleigh fading with $\AVESNR_1=\AVESNR_2=0\si{\decibel}$ and modeled by $8$-state equiprobable FSMCs. Modulation scheme is BPSK. In this case, Theorem~\ref{theo:SubOnChannel} and \ref{theo:LconvexOfImmediate} hold. Therefore, the optimal threshold $b_{1_{th}}^{*}$ and $b_{2_{th}}^{*}$ are nonincreasing in $(b_2,g_2)$ and $(b_1,g_1)$, respectively.}
	\label{fig:TH(Mono)}
\end{figure}

In Fig.~\ref{fig:TH(Mono)}, we show a monotonic optimal threshold policy when both Theorem~\ref{theo:LconvexOfQ} and Theorem~\ref{theo:SubOnChannel} hold. In this figure, $b_{i_{th}}^{*}$ is the optimal threshold defined by
\begin{equation}
    b_{i_{th}}^{*}(b_{-i},g_1,g_2)=\min\{b_i \colon a_i^*(\x)=1\}.
\end{equation}
Because of the monotonicity of $a_i^*$ in $(b_1,b_2,g_{-i})$, $b_{i_{th}}^{*}(b_{-i},g_1,g_2)$ is nonincreasing in $(b_{-i},g_{-i})$. By stacking the $b_{i_{th}}$ for all $(b_{-i},g_1,g_2)$, we can form a vector $\mathbf{b}_{th}$ with the cardinality being $|\B_{-i}|\times|\G_1|\times|\G_2|$ and convert \eqref{eq:obj} to
\begin{align}\label{eq:LConvexOpt}
        &\quad \min_{\mathbf{b}_{th}} J(\mathbf{b}_{th})   \nonumber \\
        &s.t.\ C_i(\mathbf{b}_{th})\leq{0}, \forall{i},
\end{align}
where $J$ could be the objective in \eqref{eq:obj} obtained by a Markov chain\footnote{Each $\mathbf{b}_{th}$ results in a Markov chain with the transition probability being $Pr(\x'|\x)=P_{\x\x'}^{\theta_{th}(\x)}$, where $\theta_{th}(\x)=(\IND_{\{b_1\geq b_{1_{th}}(b_2,g_1,g_2)\}},\IND_{\{b_2\geq b_{2_{th}}(b_1,g_1,g_2)\}})$.} simulation and $C_i(\mathbf{b}_{th})$ could be the constraint imposed by the monotonicity of $b_{i_{th}}$ in $(b_{-i},g_{-i})$. The method that solves \eqref{eq:LConvexOpt} could be, but not restricted to, a stochastic approximation algorithm, e.g. SPSA \cite{Huang2010}, that has lower complexity than DP and is suitable for online reinforcement learning. We will not discuss the details since it is not the purpose of our work. But, it should be clear that the results derived in this paper can be used for further study on \eqref{eq:LConvexOpt}. They are, in fact, the prerequisites for threshold policy optimization problems.

Note, that the related previous studies usually placed constraints on the environments or the DP functions in order to prove the structure in the optimal policy. For example, in \cite{Huang2010} the submodularity of the state transition probability was proved by assuming uniformly distributed traffic rates, and in \cite{Hoang2008} the strict submodularity of $Q^{(n)}$ in DP iterations was assumed to be preserved by a weight factor in the immediate cost function (However, the exact value of this factor was not given). In contrast, the results in this paper, Theorem~\ref{theo:LconvexOfImmediate}, \ref{theo:LconvexOfQ} and \ref{theo:SubOnChannel}, are essentially given in terms of unit costs and discount factor, the parameters in the MDP model. The practical meaning of them can be interpreted in two ways. If the unit costs and discount factor are adjustable, we can tune them to get a structured optimal policy. If they are given, we can check the sufficient conditions for the existence of a monotonic optimal policy after the MDP modeling. There were some results considering the uniform traffic rates, stochastic dominance of channel transition probabilities and channel modeling and modulation scheme in this paper. We either presented them as corollaries or showed the feasibility in practice by examples.

\section{Conclusion}

This paper studied an MDP modeled transmission control problem in NC-TWRC with random traffic and fading channels. The purpose was to prove the existence of a monotonic optimal transmission policy that minimized symbol delay, queue overflow, transmission power and the downlink transmission error rate in the long run. The monotonicity was established by investigating how certain properties (submodularity, $\LN$-convexity and multimodularity) of DP varied with the system parameters. This paper differed from the previous studies in two ways. Firstly, the nondecreasing optimal policy was proved to be conditioned on the parameters in the MDP model, e.g., unit costs or discount factor, instead of the constraints placed on the environment of the NC-TWRC system. Secondly, we observed the monotonicity in both queue and channel states instead of queue state only. The results in this paper can be utilized for further studies on how to simplify the optimization processes in DP, which could be useful in realtime control scenarios of NC-TWRC.

\appendices

\section{}
\label{app:SubmodularPreserve}

\subsubsection*{Proof of Lemma~\ref{lemma:SubmodularPreserve}(b)}
According to Definition~\ref{def:submodularity}, since
\begin{align}
    &\quad \sum_{i=1}^{m}\alpha_if_i(\x+\ev_v)+\sum_{i=1}^{m}\alpha_if_i(\x+\ev_w)-\sum_{i=1}^{m}\alpha_if_i(\x)-\sum_{i=1}^{m}\alpha_if_i(\x+\ev_v+\ev_w) \nonumber \\
    &=\sum_{i=1}^{m}\alpha_i\Big( f_i(\x+\ev_v)+f_i(\x+\ev_w)-f_i(\x)-f_i(\x+\ev_v+\ev_w) \Big)\geq{0}
\end{align}
for all $1\leq{v,w}\leq{n}$ such that $v\neq{w}$, so $\sum_{i=1}^{m}\alpha_if_i(\x)$ is submodular in $\x$.   \hfill\IEEEQED

\subsubsection*{Proof of Lemma~\ref{lemma:LconvexPreserve}(a)}
The case when $g$ is $\LN$-convex can be proved by Lemma $2$ and $3$\footnote{Based on Lemma $2$ and $3$ in \cite{Zipkin2008}, $f(\x)=\min_{\alpha}g(\x,\alpha)$ is $\LN$-convex in $\x$ and $\alpha^*(\x)=\arg\min_{\alpha}g(\x,\alpha)$ is nondecreasing $\x$ if $g$ is $\LN$-convex in $(\x,\alpha)$, where $\x\in\Z^n$ and $\alpha\in{\Z}$.} in \cite{Zipkin2008} through a sequential minimization, i.e., minimizing over the tuples in $\y$ one after another. The case when $g$ is multimodular can be proved by Theorem 1\footnote{$f(\x)=\min_{\alpha}g(\x,\alpha)$ is multimodular in $\x$ and $\alpha^*(\x)=\arg\min_{\alpha}g(\x,\alpha)$ is nonincreasing in $\x$ if $g$ is multimodular in $(\x,\alpha)$, where $\x\in\Z^n$ and $\alpha\in{\Z}$. } in \cite{Yu2013} in the same way.   \hfill\IEEEQED

\subsubsection*{Proof of Lemma~\ref{lemma:LconvexPreserve}(b)}
Consider the case of $\LN$-convexity first. According to Definition~\ref{def:Lconvex} and Lemma~\ref{lemma:SubmodularPreserve}(b), $\sum_{i=1}^{m}\alpha_if_i(\x-\zeta\One)$ is submodular in $(\x,\zeta)$ because of the submodularity of $f_i$ in $(\x,\zeta)$. Therefore, $\sum_{i=1}^{m}\alpha_if_i(\x)$ is $\LN$-convex.
In the same way, we can show that $\sum_{i=1}^{m}\alpha_if_i(\x)$ is multimodular if $f_i$ is multimodular for all $i$. \hfill\IEEEQED

\subsubsection*{Proof of Lemma~\ref{lemma:LconvexPreserve}(c)}
Consider function $f$ first. Since $\psi(\x,\zeta)=f(\x-\zeta\One)=h(x_1-x_2)$, according to Definition~\ref{def:Lconvex}, it suffices to show the submodularity of $h$ in $(x_1,x_2)$. But, because of the convexity of $h$,
    \begin{align}
        &\quad h(x_1+1-x_2)+h(x_1-(x_2+1))-h(x_1-x_2)-h(x_1+1-(x_2+1)) \nonumber \\
                &=h(x_1-x_2+1)+h(x_1-x_2-1)-2h(x_1-x_2)\geq{0}.
    \end{align}
By Definition~\ref{def:submodularity}, $h$ is submodular in $(x_1,x_2)$. Therefore, $f(\x)=h(x_1-x_2)$ is $\LN$-convex in $(x_1,x_2)$. Since $g(\x)=f(-M_{2,1}\x)$, according to Lemma~\ref{lemma:UniTrans}(a), $g(\x)$ is multimodular in $(x_1,x_2)$.  \hfill\IEEEQED

\section{}
\label{app:LconvexOfImmediate}

Let$i=1$ and consider the monotonicity first. $\tilde{C}(\y,\gv,\act)$ is nondecreasing in $y_1$ because
\begin{align}
    &\quad \tilde{C}(\y+\ev_1,\gv,\act)-\tilde{C}(\y,\gv,\act) \nonumber \\
        &=h_1(y_1+1)+h_2(y_2)+C_t(\gv,\act)-h_1(y_1)-h_2(y_2)-C_t(\gv,\act) \nonumber \\
        &=h_1(y_1+1)-h_1(y_1)   \nonumber \\
        &=\begin{cases}
                                \lambda>0 & -1<y_1<L_1 \\
                                \xi_o>0   & y_1=L_1 \\
                                0         & y_1=-1
                              \end{cases}.
\end{align}
$\tilde{Q}^{n}(\y,\gv,\act)$ is nondecreasing in $y_1$ because by assuming that $V^{(n-1)}(\x')$ is nondecreasing in $b'_1$
\begin{align}
    &\quad \tilde{Q}^{n}(\y+\ev_1,\gv,\act)-\tilde{Q}^{n}(\y,\gv,\act) \nonumber \\
        &=h_1(y_1+1)-h_1(y_2)+\beta \E_{\gv'} \Big[ V_\fv^{(n-1)}(\y+\ev_1,\gv')+V_\fv^{(n-1)}(\y,\gv') \Big|\gv \Big] \nonumber \\
        &=\begin{cases}
             \lambda+\beta \E_{\gv'} \bigg[ \E_\fv \Big[ V^{(n-1)}(\tilde{y}_1+1,\hat{y}_2,\gv')  \\ \qquad -V^{(n-1)}(\tilde{y}_1,\hat{y}_2,\gv') \Big] \bigg| \gv \bigg]>0 & -1<y_1<L_1 \\
             \xi_o>0   & y_1=L_1 \\
             0         & y_1=-1
          \end{cases},
\end{align}
where $\tilde{y}_1=y_1+f_1$ and $\hat{y}_2=\min\{[y_2]^+,L_2\}+f_2$.

Then, consider the multimodularity. Since the function is two-dimensional, we use the Proposition 2 in \cite{Zhuang2012} to prove the multimodularity, i.e., A function $f:\Z^2\rightarrow{\R^+}$ is multimodular if and only if it is (1) supermodualr: $\Delta_i\Delta_j f(\x)\geq{0}$ and (2) superconex: $\Delta_i f(\x+\ev_i) \geq \Delta_i f(\x+\ev_j)$ for all $i,j\in\{1,2\}$, where $\Delta_i f(\x)=f(\x)-f(\x-\ev_i)$ and $\ev_i\in{\Z^2}$ is $2$-tuple with all zero entries except the $i$th entry being one.

$\tilde{C}(\y,\gv,\act)$ is supermodular in $(y_1,a_1)$ because
\begin{align}
    &\quad \tilde{C}(\y+\ev_1,\gv,\act)+\tilde{C}(\y,\gv,\act+\ev_1)-\tilde{C}(\y,\gv,\act)-\tilde{C}(\y+\ev_1,\gv,\act+\ev_1)=0
\end{align}
and superconvex in $(y_1,a_1)$ because
\begin{align}
    &\quad \tilde{C}(\y+\ev_1,\gv,\act)-\tilde{C}(\y,\gv,\act)-\tilde{C}(\y,\gv,\act+\ev_1)+\tilde{C}(\y-\ev_1,\gv,\act+\ev_1)  \nonumber \\
        &=h_1(y_1+1)-2h_1(y_1)+h_1(y_1-1) \geq{0},
\end{align}
where the last inequality is because of the convexity of $h_1$. Therefore, $\tilde{C}(\y,\gv,\act)$ is multimodular in $(y_1,a_1)$.

By assuming the monotonicity and $\LN$-convexity of $V^{(n-1)}(\bv',\gv')$ in $b'_1$, $\tilde{Q}^{(n)}(\y,\gv,\act)$ is supermodular in $(y_1,a_1)$ because
\begin{align}
    &\quad \tilde{Q}^{n}(\y+\ev_1,\gv,\act)+\tilde{Q}^{n}(\y,\gv,\act+\ev_1)-\tilde{Q}^{n}(\y,\gv,\act)-\tilde{Q}^{n}(\y+\ev_1,\gv,\act+\ev_1)=0
\end{align}
and superconvex in $(y_1,a_1)$ because
\begin{align} \label{eq:appQueSingle4}
    &\quad \tilde{Q}^{n}(\y+\ev_1,\gv,\act)-\tilde{Q}^{n}(\y,\gv,\act)-\tilde{Q}^{n}(\y,\gv,\act+\ev_1)+\tilde{Q}^{n}(\y-\ev_1,\gv,\act+\ev_1)  \nonumber \\
        &=h_1(y_1+1)-2h_1(y_1)+h_1(y_1-1)  \nonumber \\
        &\quad +\beta \E_{\gv'} \Big[ V_\fv^{(n-1)}(\y+\ev_1,\gv')-2V_\fv^{(n-1)}(\y,\gv')+V_\fv^{(n-1)}(\y-\ev_1,\gv') \Big|\gv \Big] \nonumber \\
        &=\begin{cases}
             \beta \E_{\gv'} \bigg[ \E_\fv \Big[ V^{(n-1)}(\tilde{y}_1+1,\hat{y}_2,\gv')\\
             \qquad\qquad -2V^{(n-1)}(\tilde{y}_1,\hat{y}_2,\gv')+V^{(n-1)}(\tilde{y}_1-1,\hat{y}_2,\gv') \Big] \bigg| \gv \bigg] \geq 0 & 0<y_1<L_1\\
             \lambda+\beta \E_{\gv'} \bigg[ \E_\fv \Big[ V^{(n-1)}(1+f_1,\hat{y}_2,\gv')  \\ \qquad\qquad -V^{(n-1)}(f_1,\hat{y}_2,\gv') \Big] \bigg| \gv \bigg] \geq 0 & y_1=0 \\
		     \xi_o-\lambda+\beta \E_{\gv'} \bigg[ \E_\fv \Big[ -V^{(n-1)}(L_1+f_1,\hat{y}_2,\gv')  \\
             \qquad\qquad +V^{(n-1)}(L_1-1+f_1,\hat{y}_2,\gv') \Big] \bigg| \gv \bigg] \geq 0 & y_1=L_1
          \end{cases},
\end{align}
where $\tilde{y}_1=y_1+f_1$ and $\hat{y}_2=\min\{[y_2]^+,L_2\}+f_2$. The inequality in \eqref{eq:appQueSingle4} in the case of $0<y_1<L_1$ is because of the $\LN$-convexity of $V^{(n-1)}(\bv',\gv')$ in $b'_1$ and Lemma~\ref{lemma:LconvexPreserve}(d). The inequality in the case of $y_1=0$ is because of the monotonicity of $V^{(n-1)}(\bv',\gv')$ in $b'_1$. The inequality in the case of $y_1=L_1$ is explained as follows.

Denote $\act^*(\x')=\arg\min_{\act'}Q^{(n-1)}(\x',\act')$ and recall that $V^{(n-1)}(\x')=Q^{(n-1)}(\x',\act^*(\x'))$. Due to the $\LN$-convexity of $Q^{(n)}(\x',\act')$ in $(b'_1,a'_1)$, $a_1^*(b'_1,b'_2,\gv) \geq a_1^*(b'_1-1,b'_2,\gv')$. Because of the monotonicity of $V^{(n-2)}$ in $b''_1$, it is easy to show that the lower bound of $-V^{(n-1)}(b'_1,b'_2,\gv')+V^{(n-1)}(b'_1-1,b'_2,\gv')$ is given by when $\act^*(b'_1,b'_2,\gv')=(1,1)$ and $\act^*(b'_1-1,b'_2,\gv')=(0,0)$, i.e., we have
\begin{align} \label{eq:appQueSingle5}
    &\quad -V^{(n-1)}(b'_1,b'_2,\gv')+V^{(n-1)}(b'_1-1,b'_2,\gv')  \nonumber \\
        &=-Q^{(n-1)}(b'_1,b'_2,\gv',\act^*(b'_1,b'_2,\gv'))+Q^{(n-1)}(b'_1-1,b'_2,\gv',\act^*(b'_1-1,b'_2,\gv')) \nonumber \\
        &\geq -Q^{(n-1)}(b'_1,b'_2,\gv',(1,1))+Q^{(n-1)}(b'_1-1,b'_2,\gv',(0,0)) \nonumber \\
        &\geq -C(b'_1,b'_2,\gv',(1,1))+C(b'_1-1,b'_2,\gv',(0,0)) \nonumber \\
        &\geq -\lambda-\eta(P_e(g'_1)+P_e(g'_2))-\tau \nonumber \\
        &\geq -\lambda-\eta-\tau,
\end{align}
where last inequality is by knowing $P_e(g'_i)\leq{0.5}, \forall{i}\in\{1,2\}$. Since $\xi_o \geq{2\lambda+\eta+\tau}$, we have the inequality in the case of $y_1=L_1$ in \eqref{eq:appQueSingle4}.

Therefore, $\tilde{Q}^{(n)}(\y,\gv,\act)$ is $\LN$-convex in $(y_1,a_1)$.

In the case of $i=2$, the $\LN$-convexity of $\tilde{C}(\y,\gv,\act)$ and $\tilde{Q}^{(n)}(\y,\gv,\act)$ in $(y_2,a_2)$ can be prove in the same way.   \hfill\IEEEQED

\section{}
\label{app:TeamGame}
$Q^{(n)}(\x,\act)$ is strictly submodular in $\act$ because
\begin{align} \label{eq:TeamGameS}
    &\quad Q^{(n)}(\x,(1,0))+Q^{(n)}(\x,(0,1))-Q^{(n)}(\x,(0,0))-Q^{(n)}(\x,(1,1))  \nonumber \\
        &=\tau+\beta \E_{\gv'} \Big[ V_\fv^{(n-1)}(b_1-1,b_2,\gv')+V_\fv^{(n-1)}(b_1,b_2-1,\gv') \nonumber \\
        &\qquad -V_\fv^{(n-1)}(b_1,b_2,\gv')-V_\fv^{(n-1)}(b_1-1,b_2-1,\gv') \Big|\gv \Big] \nonumber \\
        &= \begin{cases}		
		      \tau+\beta \E_{\gv'} \Big[ \E_\fv[ V^{(n-1)}(\tilde{b}_1-1,\tilde{b}_2,\gv')\\
                \quad +V^{(n-1)}(\tilde{b}_1,\tilde{b}_2-1,\gv')-V^{(n-1)}(\tilde{b}_1,\tilde{b}_2,\gv')\\
                \qquad -V^{(n-1)}(\tilde{b}_1-1,\tilde{b}_2-1,\gv') ]\Big|\gv \Big]>0 & -1<b_i-a_i<L_i+1,\forall{i}\in{\{1,2\}}\\
		      \tau>0 & otherwise
	       \end{cases},
\end{align}
where $\tilde{b}_i=b_i+f_i$. In \eqref{eq:TeamGameS}, the inequality in the case of $-1<b_i-a_i<L_i+1,\forall{i}\in{\{1,2\}}$ is because of the $\LN$-convexity of $V^{(n-1)}(\bv',\gv')$ in $\bv'$. Therefore, $-Q^{(n)}(\x,\act)$ is strictly supermodular in $\act$. According to the definition in \cite{Milgrom1990}, the game is supermodular.
  \hfill\IEEEQED

\section{}
\label{app:LconvexOfQ}

$C_{h}(\bv,\act)$ is $\LN$-convex in $(\bv,\act)$ because of the convexity of $h_i$, and $C_{t}(\gv,\act)$ is $\LN$-convex in $(\bv,\act)$ because
\begin{align} \label{eq:appImmLconvex1}
    &\quad C_{t}(\gv,(1,0))+C_{t}(\gv,(0,1))-C_{t}(\gv,(0,0))-C_{t}(\gv,(1,1))  \nonumber \\
        &=\eta P_e(g_2)+\tau+\eta P_e(g_1)+\tau-\eta(P_e(g_1)+P_e(g_2))-\tau \nonumber \\
        &=\tau >0.
\end{align}
Note, since $\act\in\{0,1\}^2$, according to Proposition 2\footnote{A function $f\colon\Z^2\rightarrow{\R}$ is $\LN$-convex if and only if it is (1) submodular: $\Delta_i\Delta_j f(\x)\leq{0}$ and (2) subconex: $\Delta_i f(\x+\ev_i) \geq \Delta_i f(\x-\ev_j)$ for all $i,j\in\{1,2\}$, where $\Delta_i f(\x)=f(\x)-f(\x-\ev_i)$ and $\ev_i\in{\Z^2}$ is $2$-tuple with all zero entries except the $i$th entry being one.} in \cite{Zhuang2012}, the submodularity by \eqref{eq:appImmLconvex1} is sufficient to show the $\LN$-convexity of $C_{t}(\gv,\act)$ in $(\bv,\act)$. Therefore, by Lemma~\ref{lemma:LconvexPreserve}(b), $C$ is $\LN$-convex in $(\bv,\act)$.

Then, consider the $\LN$-convexity of $Q$ in $(\bv,\act)$. Let $BR_i(a_{-i})=\arg\min_{a_{i}}Q^{(n)}(\x,(a_i,a_{-i}))$. Equilibria $(0,0),(1,1)$ implies $BR_i(a_{-i})=a_{-i}$, i.e., $a_1=a_2$. Consider
\begin{align} \label{eq:appLconvexOfQ1}
    &\quad Q^{(n)}(\bv,\gv,(a_1,a_1))   \nonumber \\
    &=\tilde{C}_h(\bv-a_1\One)+a_1 \bigg( \eta(P_e(g_1)+P_e(g_2))+\tau \bigg) +\beta \E_{\gv'} \Big[ V_\fv^{(n-1)}(\bv-a_1\One,\gv') \Big|\gv \Big].
\end{align}
$Q^{(n)}(\x,(a_1,a_1))$ is $\LN$-convex in $(\bv,a_1)$ since: When $b_i-a_1<L_i+1$ for all ${i\in\{1,2\}}$, $Q^{(n)}$ is $\LN$-convex in $(\bv,a_1)$ because of the $\LN$-convexity of $V^{(n-1)}$ in $\bv'$ and Lemma~\ref{lemma:LconvexPreserve}(d) and (e); When $b_i-a_1=L_i+1$ for either $i=1$ or $i=2$, the $\LN$-convexity of $Q^{(n)}$ can be shown in the same way as in Appendix~\ref{app:LconvexOfImmediate} under condition $\xi_o\geq{2\lambda+\tau+\eta}$. By Theorem~\ref{theo:PstarProperty} and Proposition~\ref{prop:PropertyDP}, $V^*(\x)$ is $\LN$-convex in $\bv$ and the optimal action $\act^*$ is nondecreasing in $\bv$.     \hfill\IEEEQED

\section{}
\label{app:BREquiProb}
We just need to show that condition (b) in Theorem~\ref{theo:LconvexOfQ} is satisfied. Let $b_i-a_i<L_i+1$ for all $i\in\{1,2\}$. It suffices to show $BR_i(a_{-i})=a_{-i}$ for all $i\in\{1,2\}$ in order to prove equilibria $(0,0),(1,1)$ in Theorem~\ref{theo:LconvexOfQ}. Because the game has strictly supermodular utility, $BR_i(a_{-i}+1)>BR_i(a_{-i})$. So $BR_i(1)=1$, if we can prove $BR_i(0)=0$. By knowing that $p_1=0.5$, we can show that
\begin{align}
    &\quad Q^{(n)}(\bv,\gv,(1,0))-Q^{(n)}(\bv,\gv,(0,0)) \nonumber \\
    &=\begin{cases}
          -\lambda+\tau+\eta P_e(g_1) \\
          \quad+0.5\beta \Big( V(b_1-1,\hat{b}_2),\gv')-V(b_1+1,\hat{b}_2),\gv')\Big) \geq{0}  & 0<b_1<L_1+1 \\
          -\lambda+\tau+\eta P_e(g_1) \geq{0}  & \text{otherwise}
      \end{cases},
\end{align}
where $\hat{b}_2=\min\{[b_2]^+,L_2\}+f_2$ and the inequality in the case when $0<b_1<L_1+1$ is because that, by a similar approach as in \eqref{eq:appQueSingle5}, we can show that
\begin{align} \label{eq:appBREquiProb2}
    &\quad V^{(n-1)}(b'_1-1,b'_2,\gv')-V^{(n-1)}(b'_1+1,b'_2,\gv') \nonumber \\
    &=Q^{(n-1)}(b'_1-1,b'_2,\gv',\act^*(b'_1-1,b'_2,\gv'))-Q^{(n-1)}(b'_1+1,b'_2,\gv',\act^*(b'_1+1,b'_2,\gv')) \nonumber \\
    &\geq  C(b'_1-1,b'_2,\gv',(0,0))-C(b'_1+1,b'_2,\gv',(1,1))  \nonumber \\
    &= -\tau-\eta(P_e(g'_1)+P_e(g'_2)) \nonumber \\
    &\geq  -\tau-\eta
\end{align}
and the condition $\beta\leq\frac{2(\tau-\lambda)}{\tau+\eta}$.

Similarly, we can show $Q^{(n)}(\bv,\gv,(0,1))-Q^{(n)}(\bv,\gv,(0,0))\geq{0}$ in the case when $p_2=0.5$. So, $BR_i(a_{-i})=a_{-i}$. \hfill\IEEEQED

\section{}
\label{app:SubOnChannel}
Let $i=2$. $C(\x,\act)$ is submodular in $(b_2,g_1,a_2)$ because
\begin{align} \label{eq:appChan1}
    &\quad C(\bv+\ev_2,\gv,\act)+C(\bv,\gv,\act+\ev_2)-C(\bv,\gv,\act)-C(\bv+\ev_2,\gv,\act+\ev_2)  \nonumber \\
        &=h_2(b_2+1-a_2)+h_2(b_2-a_2-1)-2h_2(b_2-a_2)\geq{0},
\end{align}
\begin{align} \label{eq:appChan2}
    &\quad C(\bv+\ev_2,\gv,\act)+C(\bv,\gv+\ev_1,\act)-C(\bv,\gv,\act)-C(\bv+\ev_2,\gv+\ev_1,\act)  \nonumber \\
        &=h_2(b_2+1-a_2)+C_t(\gv,\act)+h_2(b_2-a_2)+C_t(\gv+\ev_1,\act)  \nonumber \\
        &\qquad  -h_2(b_2-a_2)-C_t(\gv,\act)-h_2(b_2+1-a_2)-C_t(\gv+\ev_1,\act)=0,
\end{align}
\begin{align} \label{eq:appChan3}
    &\quad C(\bv,\gv,\act+\ev_2)+C(\bv,\gv+\ev_1,\act)-C(\bv,\gv,\act)-C(\bv,\gv+\ev_1,\act+\ev_2)  \nonumber \\
        &=\eta (a_2+1) Pe(g_1)+\eta a_2 Pe(g_1+1)-\eta a_2 Pe(g_1)-\eta(a_2+1) Pe(g_1+1)   \nonumber \\
        &=\eta (P_e(g_1)-P_e(g_1+1))\geq{0}.
\end{align}

By assuming the submodularity of $V^{(n-1)}(\bv',\gv')$ in $(b'_2,g'_1)$, $Q^{(n)}(\x,\act)$ is submodular in $(b_2,g_1,a_2)$ because
\begin{align} \label{eq:appChan4}
    &\quad Q^{(n)}(\bv+\ev_2,\gv,\act)+Q^{(n)}(\bv,\gv,\act+\ev_2)-Q^{(n)}(\bv,\gv,\act)-Q^{(n)}(\bv+\ev_2,\gv,\act+\ev_2)  \nonumber \\
        &=h_2(b_2+1-a_2)+h_2(b_2-a_2-1)-2h_2(b_2-a_2) \nonumber \\
        &\qquad +\beta \E_{\gv'} \Big[ V_\fv^{(n-1)}(b_1-a_1,b_2+1-a_2,\gv')+V_\fv^{(n-1)}(b_1-a_1,b_2-a_2-1,\gv') \nonumber \\
        &\qquad\qquad  -2V_\fv^{(n-1)}(b_1-a_1,b_2-a_2,\gv') \Big|\gv \Big] \geq{0},
\end{align}
\begin{align} \label{eq:appChan5}
    &\quad Q^{(n)}(\bv+\ev_2,\gv,\act)+Q^{(n)}(\bv,\gv+\ev_1,\act)-Q^{(n)}(\bv,\gv,\act)-Q^{(n)}(\bv+\ev_2,\gv+\ev_1,\act)  \nonumber \\
        &= \beta \bigg( \E_{\gv'} \Big[ V_\fv^{(n-1)}(b_1-a_1,b_2+1-a_2,\gv') \nonumber \\
        &\qquad -V_\fv^{(n-1)}(b_1-a_1,b_2-a_2,\gv') \Big|\gv \Big] - \E_{(\gv+\ev_1)'} \Big[ V_\fv^{(n-1)}(b_1-a_1,b_2+1-a_2,\gv') \nonumber \\
        &\qquad\qquad -V_\fv^{(n-1)}(b_1-a_1,b_2-a_2,(\gv+\ev_1)') \Big|(\gv+\ev_1) \Big] \bigg) \geq{0},
\end{align}
\begin{align} \label{eq:appChan6}
    &\quad Q^{(n)}(\bv,\gv,\act+\ev_2)+Q^{(n)}(\bv,\gv+\ev_1,\act)-Q^{(n)}(\bv,\gv,\act)-Q^{(n)}(\bv,\gv+\ev_1,\act+\ev_2)  \nonumber \\
        &=\eta (P_e(g_1)-P_e(g_1+1))+ \beta \E_{\gv'} \Big[ V_\fv^{(n-1)}(b_1-a_1,b_2-1,\gv') \nonumber \\
        &\qquad -V_\fv^{(n-1)}(b_1-a_1,b_2,\gv') \Big|\gv \Big] - \E_{(\gv+\ev_1)'} \Big[ V_\fv^{(n-1)}(b_1-a_1,b_2-1,(\gv+\ev_1)') \nonumber \\
        &\qquad\qquad  -V_\fv^{(n-1)}(b_1-a_1,b_2,(\gv+\ev_1)') \Big|(\gv+\ev_1) \Big] \geq{0}.
\end{align}
The inequality in \eqref{eq:appChan4} is because of the convexity of $h_2$ and the $\LN$-convexity of $V^{(n-1)}(\bv',\gv')$ in $b'_2$ under condition $\xi_o\geq 2\lambda+\eta+\tau$. The inequality in \eqref{eq:appChan5} is because of the submodularity of $V^{(n-1)}(\bv',\gv')$ in $(b'_2,g'_1)$ and the first order stochastic monotonicity of $\Pggi$ in $g_i$. The last inequality in \eqref{eq:appChan6} is explained as follows.

By using a similar approach as in \eqref{eq:appQueSingle5}, we can show that
\begin{align} \label{eq:appChan7}
    &\quad V^{(n-1)}(\bv'-\ev_2,\gv')+V^{(n-1)}(\bv',\gv'+\ev_1)-V^{(n-1)}(\bv',\gv')-V^{(n-1)}(\bv'-\ev_2,\gv'+\ev_1)  \nonumber \\
        &=Q^{(n-1)}(\bv'-\ev_2,\gv',\act^*(\bv'-\ev_2,\gv'))+Q^{(n-1)}(\bv',\gv'+\ev_1,\act^*(\bv',\gv'+\ev_1)) \nonumber \\
        &\qquad -Q^{(n-1)}(\bv',\gv',\act^*(\bv',\gv'))-Q^{(n-1)}(\bv'-\ev_2,\gv'+\ev_1,\act*(\bv'-\ev_2,\gv'+\ev_1)) \nonumber \\
        &\geq Q^{(n-1)}(\bv'-\ev_2,\gv',\act^*(\bv'-\ev_2,\gv'))+Q^{(n-1)}(\bv',\gv'+\ev_1,\act^*(\bv',\gv'+\ev_1)) \nonumber \\
        &\qquad -Q^{(n-1)}(\bv',\gv',\act^*(\bv',\gv'+\ev_1))-Q^{(n-1)}(\bv'-\ev_2,\gv'+\ev_1,\act^*(\bv'-\ev_2,\gv')) \nonumber \\
        &\geq Q^{(n-1)}(\bv'-\ev_2,\gv',(0,0))+Q^{(n-1)}(\bv',\gv'+\ev_1,(1,1)) \nonumber \\
        &\qquad -Q^{(n-1)}(\bv',\gv',(1,1))-Q^{(n-1)}(\bv'-\ev_2,\gv'+\ev_1,(0,0)) \nonumber \\
        &\geq C(\bv'-\ev_2,\gv',(0,0))+C(\bv',\gv'+\ev_1,(1,1))  \nonumber \\
        &\qquad -C(\bv',\gv',(1,1))-C(\bv'-\ev_2,\gv'+\ev_1,(0,0)) \nonumber \\
        &=\eta (P_e(g'_1+1)-P_e(g'_1)).
\end{align}
Because of the condition $\beta\leq\frac{P_e(g_i)-P_e(g_i+1)}{\sum_{g'_i}P_{g_ig'_i} ( P_e(g'_i)-P_e(g'_i+1) ) }$, we have the inequality in \eqref{eq:appChan6}.

Similarly, we can show that Theorem~\ref{theo:SubOnChannel} holds for $i=1$. Therefore, by Proposition~\ref{prop:PropertyDP} and Theorem~\ref{theo:PstarProperty}, $V^{*}(\x)$ is submodular in $(b_{i},g_{-i})$ and the optimal action $a_i^*$ is nondecreasing in $(b_{i},g_{-i})$.  \hfill\IEEEQED

\section{}
\label{app:EquiFSMC}

In an equiprobable partition Rayleigh fading FSMC, $P_e(g_i)$ is nonincreasing in $g_i$, i.e., $P_e(g_i) \geq P_e(g_i+1)$. Because of slow and flat fading assumption, the channel transitions can be worked out by level crossing rate (LCR) \cite{Sadeghi2008} and only happens between adjacent states, i.e., $g'_i\in\{g_i-1, g_i, g_i+1\}$. Further, $P_{gg'}=P_{g'g}$, and $P_{gg'}\ll{P_{gg}}$ for all $g'\neq{g}$. According to Definition~\ref{def:1stStoDom}, for nondecreasing $u$, $\Pggi$ is first order stochastic nondecreasing in $g_i$ because
\begin{align}
        &\quad \sum_{(g_i+1)'}P_{(g_i+1)(g_i+1)'}u \Big( (g_i+1)' \Big)-\sum_{g'_i}P_{g_ig'_i}u(g'_i) \nonumber \\
        &\geq (1-2P_{g_ig_i+1}) \Big( u(g_i+1)-u(g_i) \Big) \geq 0,
\end{align}
where $1-2P_{g_ig_i+1}\geq{0}$ is because $P_{gg'}\ll{P_{gg}}$ and $\sum_{g'}P_{gg'}=1$.   \hfill\IEEEQED

\bibliography{IEEEabrv,ding-arxiv}

\begin{thebibliography}{10}
\providecommand{\url}[1]{#1}
\csname url@samestyle\endcsname
\providecommand{\newblock}{\relax}
\providecommand{\bibinfo}[2]{#2}
\providecommand{\BIBentrySTDinterwordspacing}{\spaceskip=0pt\relax}
\providecommand{\BIBentryALTinterwordstretchfactor}{4}
\providecommand{\BIBentryALTinterwordspacing}{\spaceskip=\fontdimen2\font plus
\BIBentryALTinterwordstretchfactor\fontdimen3\font minus
  \fontdimen4\font\relax}
\providecommand{\BIBforeignlanguage}[2]{{%
\expandafter\ifx\csname l@#1\endcsname\relax
\typeout{** WARNING: IEEEtran.bst: No hyphenation pattern has been}%
\typeout{** loaded for the language `#1'. Using the pattern for}%
\typeout{** the default language instead.}%
\else
\language=\csname l@#1\endcsname
\fi
#2}}
\providecommand{\BIBdecl}{\relax}
\BIBdecl

\bibitem{Ahlswede2000}
R.~Ahlswede, N.~Cai, S.-Y.~R. Li, and R.~W. Yeung, ``Network information
  flow,'' \emph{{IEEE} Trans. Inf. Theory}, vol.~46, no.~4, pp. 1204--1216,
  Jul. 2000.

\bibitem{WuXOR2004}
Y.~Wu, ``Information exchange in wireless networks with network coding and
  physical-layer broadcast,'' Microsoft Research, Redmond WA, Tech. Rep.
  MSR-TR-2004-78, Aug. 2004.

\bibitem{KattiXOR2006}
S.~Katti, H.~Rahul, W.~Hu, D.~Katabi, M.~M{\'e}dard, and J.~Crowcroft, ``Xors
  in the air: practical wireless network coding,'' \emph{SIGCOMM Comput.
  Commun. Rev.}, vol.~36, no.~4, pp. 243--254, Aug. 2006.

\bibitem{Hausl2006}
C.~Hausl and J.~Hagenauer, ``Iterative network and channel decoding for the
  two-way relay channel,'' in \emph{Proc. IEEE Int. Conf. on Commun. 2006},
  Istanbul, 2006, pp. 1568--1573.

\bibitem{Gao2009}
F.~Gao, R.~Zhang, and Y.-C. Liang, ``Optimal channel estimation and training
  design for two-way relay networks,'' \emph{{IEEE} Trans. Commun.}, vol.~57,
  no.~10, pp. 3024--3033, Oct. 2009.

\bibitem{Koike2009}
T.~Koike-Akino, P.~Popovski, and V.~Tarokh, ``Optimized constellations for
  two-way wireless relaying with physical network coding,'' \emph{{IEEE} J.
  Sel. Areas Commun.}, vol.~27, no.~5, pp. 773--787, Jun. 2009.

\bibitem{KattiImportance2005}
S.~Katti, D.~Katabi, W.~Hu, H.~Rahul, and M.~Medard, ``The importance of being
  opportunistic: Practical network coding for wireless environments,'' in
  \emph{Proc. 43rd Annu. Allerton Conf. on Commun., Control, and Computing},
  Monticello, IL, 2005, pp. 756--765.

\bibitem{ChenONC2007}
W.~Chen, K.~Letaief, and Z.~Cao, ``Opportunistic network coding for wireless
  networks,'' in \emph{Proc. IEEE Int. Conf. on Commun. 2007}, Glasgow, 2007,
  pp. 4634 --4639.

\bibitem{HsuONC2011}
Y.-P. Hsu, N.~Abedini, S.~Ramasamy, N.~Gautam, A.~Sprintson, and S.~Shakkottai,
  ``Opportunities for network coding: To wait or not to wait,'' in \emph{Proc.
  IEEE Int. Symp. Inform. Theory}, St. Petersburg, 2011, pp. 791 --795.

\bibitem{Ding2012}
N.~Ding, I.~Nevat, G.~W. Peters, and J.~Yuan, ``Opportunistic network coding
  for two-way relay fading channels,'' presented at IEEE Int. Conf. on Commun.
  2013, Budapest, 2013.

\bibitem{Sadeghi2008}
P.~Sadeghi, R.~A. Kennedy, P.~B. Rapajic, and R.~Shams, ``Finite-state {M}arkov
  modeling of fading channels: A survey of principles and applications,''
  \emph{{IEEE} Signal Process. Mag.}, vol.~25, no.~5, pp. 57--80, Sep. 2008.

\bibitem{SuttonRL1998}
R.~S. Sutton and A.~G. Barto, \emph{Introduction to Reinforcement Learning},
  1st~ed.\hskip 1em plus 0.5em minus 0.4em\relax Cambridge, MA: MIT Press,
  1998.

\bibitem{Smith2002}
J.~E. Smith and K.~F. McCardle, ``Structural properties of stochastic dynamic
  programs,'' \emph{Operations Research}, vol.~50, no.~5, pp. 796--809,
  Sep./Oct. 2002.

\bibitem{Djonin2007}
D.~V. Djonin and V.~Krishnamurthy, ``{MIMO} transmission control in fading
  channels--a constrained {M}arkov decision process formulation with monotone
  randomized policies,'' \emph{{IEEE} Trans. Signal Process.}, vol.~55, no.~10,
  pp. 5069--5083, Oct. 2007.

\bibitem{Huang2010}
J.~Huang and V.~Krishnamurthy, ``Transmission control in cognitive radio as a
  {M}arkovian dynamic game: Structural result on randomized threshold
  policies,'' \emph{{IEEE} Trans. Commun.}, vol.~58, no.~1, pp. 301--310, Jan.
  2010.

\bibitem{Topkis2001}
D.~M. Topkis, \emph{Supermodularity and complementarity}.\hskip 1em plus 0.5em
  minus 0.4em\relax Princeton: Princeton University Press, 2001.

\bibitem{Murota2005}
K.~Murota, ``Note on multimodularity and {L}-convexity,'' \emph{Math. of
  Operations Research}, vol.~30, no.~3, pp. 658--661, Aug. 2005.

\bibitem{Zhuang2012}
W.~Zhuang and M.~Z. Li, ``Monotone optimal control for a class of {M}arkov
  decision processes,'' \emph{European J. of Operational Research}, vol. 217,
  no.~2, pp. 342 -- 350, Mar. 2012.

\bibitem{Zipkin2008}
P.~Zipkin, ``On the structure of lost-sales inventory models.''
  \emph{Operations research}, vol.~58, no.~4, pp. 937--944, Jul./Aug. 2008.

\bibitem{Pang2012}
Z.~Pang, F.~Y. Chen, and Y.~Feng, ``Technical note--a note on the structure of
  joint inventory-pricing control with leadtimes,'' \emph{Operations Research},
  vol.~60, no.~3, pp. 581--587, May/Jun. 2012.

\bibitem{Hoang2008}
A.~T. Hoang and M.~Motani, ``Cross-layer adaptive transmission: Optimal
  strategies in fading channels,'' \emph{{IEEE} Trans. Commun.}, vol.~56,
  no.~5, pp. 799--807, May 2008.

\bibitem{Yu2013}
Q.~L.~P. Yu, ``Multimodularity and structural properties of stochastic dynamic
  programs,'' \emph{Working Paper. School of Bus. and Manage., HongKong
  University of Sci. and Tech.}, 2013.

\bibitem{Yang2012}
L.~Yang, Y.~E. Sagduyu, and J.~H. Li, ``Adaptive network coding for scheduling
  real-time traffic with hard deadlines,'' in \emph{Proc. of 13th ACM Int.
  Symp. on Mobile Ad Hoc Networking and Computing}, New York, 2012, pp.
  105--114.

\bibitem{PutermanMDP1994}
M.~L. Puterman, \emph{Markov Decision Processes: Discrete Stochastic Dynamic
  Programming}, 1st~ed.\hskip 1em plus 0.5em minus 0.4em\relax New York: John
  Wiley \& Sons, Inc., 1994.

\bibitem{Topkis1978}
D.~M. Topkis, ``Minimizing a submodular function on a lattice,''
  \emph{Operations Research}, vol.~26, no.~2, pp. 305--321, Mar./Apr. 1978.

\bibitem{Hajek1985}
B.~Hajek, ``Extremal splittings of point processes,'' \emph{Math. of Operations
  Research}, vol.~10, no.~4, pp. 543--556, Nov. 1985.

\bibitem{Murota2003}
K.~Murota, \emph{Discrete convex analysis}.\hskip 1em plus 0.5em minus
  0.4em\relax Philadelphia: SIAM, 2003.

\bibitem{Altman2000}
E.~Altman, B.~Gaujal, and A.~Hordijk, ``Multimodularity, convexity, and
  optimization properties,'' \emph{Math. of Operations Research}, vol.~25,
  no.~2, pp. 324--347, May 2000.

\bibitem{Milgrom1990}
P.~Milgrom and J.~Roberts, ``Rationalizability, learning, and equilibrium in
  games with strategic complementarities,'' \emph{Econometrica: J. of the
  Econometric Society}, vol.~58, no.~6, pp. 1255--1277, Nov. 1990.

\end{thebibliography}

\end{document}